%% file: ms.tex
\documentclass[numbers, review, 12pt]{elsarticle}

\usepackage{graphicx}

\usepackage{amssymb}
\usepackage[hidelinks]{hyperref}

\graphicspath{{figures/}}

\usepackage{tensorNotation}
\usepackage{algpseudocode}
\usepackage{algorithm}
\usepackage{setspace}
\usepackage{pgfplots}
\usepackage{caption}
\usepackage[size=footnotesize]{subcaption}
\usepackage{amsmath}
\usepackage{etoolbox}
\usepackage{cleveref}
\usepackage[textsize=tiny, textwidth=1.8cm]{todonotes}
\usepackage{everysel}

\usepackage{tensorNotation}

\journal{Computers \& Fluids}

\usepackage{longtable}
\usepackage{multirow}
\usepackage{booktabs}
\usepackage{cleveref}
\usepackage{everypage}
\usepackage{mathtools}
\usepackage{geometry}

\usepackage[T1]{fontenc} 

\usepackage[amssymb]{SIunits}

\usepackage{tikz}
\usetikzlibrary{positioning,shapes,arrows,chains}
\makeatletter
\let\tikz@preaction@layer=\pgfutil@empty     
\tikzset{preaction layer/.store in=\tikz@preaction@layer} 
\makeatother

\tikzstyle{rounded shadow} = [
    copy shadow={%
      preaction layer=shadow,
      fill=gray!25,
      draw=none,
      shadow xshift=0.2em,
      shadow yshift=-0.25em
   }]

\tikzstyle{test shadow} = [
    copy shadow={%
      preaction layer=shadow,
      fill=gray!25,
      draw=none,
      shadow xshift=0.1em,
      shadow yshift=-0.2em
   }]

\pgfdeclarelayer{shadow} 
\pgfsetlayers{shadow,main}
\usepackage{xcolor}
\begin{document}


\newcommand{\ie}{i.e.\ }
\newcommand{\wrt}{w.r.\ to\ }

\newcommand{\Fsigma}{\ensuremath{\F{\Sigma}}}
\renewcommand{\U}{\ensuremath{\mathbf{v}}}
\DeclareRobustCommand{\rchitild}{{\mathpalette\tirchi\relax}}
\DeclareRobustCommand{\rchi}{{\mathpalette\irchi\relax}}
\newcommand{\irchi}[2]{\raisebox{\depth}{$#1\chi$}} 
\newcommand{\tirchi}[2]{\raisebox{\depth}{$#1\tilde{\chi}$}} 

\newcommand{\CellNeighborhood}{\ensuremath{\mathcal{C}}}
\newcommand{\CellMap}{\ensuremath{\mathcal{M}_{c} }}
\newcommand{\cf}{\ensuremath{\mathbf{cf}}}
\renewcommand{\div}[1]{\ensuremath{\nabla\cdot#1}}
\newcommand{\Front}{\ensuremath{\Gamma}}
\renewcommand{\laplacian}[1]{\ensuremath{\nabla\cdot(\nabla #1)}}
\newcommand{\Neighborhood}{\ensuremath{\mathcal{N}}}
\newcommand{\Nsigma}{\ensuremath{\mathbf{n}_{\Sigma}}}
\newcommand{\Oh}{\ensuremath{\mathcal{O}}}
\newcommand{\ft}{\ensuremath{\tilde{f}}}
\newcommand{\Sf}{\ensuremath{\mathbf{S}_f}}
\newcommand{\Sft}{\ensuremath{\mathbf{S}_{\tf}}}
\newcommand{\Sfhat}{\ensuremath{\mathbf{\hat{S}}_f}}
\newcommand{\Sfhatt}{\ensuremath{\mathbf{\hat{S}}_{\tf}}}
\newcommand{\Sngrad}{\ensuremath{\nabla_n}}
\newcommand{\Point}{\ensuremath{\vec{p}}}
\newcommand{\phib}{\ensuremath{\boldsymbol{\phi}}}
\newcommand{\Vertex}{\ensuremath{\vec{v}}}
\newcommand{\Triangle}{\ensuremath{\mathcal{T}}}
\newcommand{\Trianglei}[1]{\ensuremath{\Triangle_{#1}}}
\newcommand{\TriangleOther}{\ensuremath{\Triangle_\CellOther}}
\newcommand{\TriangleNearest}{\ensuremath{\Triangle_N}}
\newcommand{\TriangleVertex}[1]{\ensuremath{\vec{v}_{\Triangle, #1}}}
\newcommand{\TriangleNormal}{\ensuremath{\vec{n}_\Triangle}}
\newcommand{\TriangleNormalHat}{\ensuremath{\hat{\vec{n}}_\Triangle}}
\newcommand{\TriangleNormalSum}{\ensuremath{\vec{n}_{\Triangle,sum}}}
\newcommand{\TriangleNormalHatSum}{\ensuremath{\hat{\mathbf{n}}_{\Triangle,sum}}}
\newcommand{\TriangleNormali}[1]{\ensuremath{\vec{n}_{\Triangle,#1}}}
\newcommand{\TriangleNormalOther}{\TriangleNormali{\CellOther}}
\newcommand{\Triangulation}{\ensuremath{T}}
\newcommand{\TriangulationPoint}{\ensuremath{\Point_T}}
\newcommand{\TriangleIntersection}{\ensuremath{\mathbf{x}_{\Triangle_N}}}
\newcommand{\TriangleCells}{\ensuremath{\CellMap(\Triangle)}} 
\newcommand{\Tetrahedron}{\ensuremath{\mathcal{T}_e}}
\newcommand{\TetraVolumei}

\newcommand{\F}[1]{\ensuremath{\mathbf{f}_#1}}
\newcommand{\Fsigmac}{\ensuremath{\F{\Sigma,c}}}
\newcommand{\Finterpolate}{\ensuremath{F}}
\newcommand{\Reconstruct}{\ensuremath{R}}

\newcommand{\Domain}{\ensuremath{\Omega}}
\newcommand{\Boundary}{\ensuremath{\partial\Domain}}
\newcommand{\Interface}{\ensuremath{\Sigma}}
\newcommand{\MeanCurvature}{\ensuremath{\kappa}}
\newcommand{\ExactCurvature}{\ensuremath{\MeanCurvature_\text{exact}}}
\newcommand{\SignedDistance}{\ensuremath{\phi}}
\newcommand{\PhaseIndicator}{\ensuremath{\alpha}}
\newcommand{\Cell}{\ensuremath{c}}
\newcommand{\Face}{\ensuremath{f}}
\newcommand{\TriangleMap}{\ensuremath{\mathcal{M}_\Triangle}}
\newcommand{\Semiaxes}{\ensuremath{\mathbf{s}}}
\newcommand{\LBop}{\ensuremath{\Delta_\Interface}}
\newcommand{\Min}{\text{min}}
\newcommand{\Max}{\text{max}}
\newcommand{\Massflux}{\ensuremath{\dot{m}_\Face}}
\newcommand{\Volflux}{\ensuremath{\dot{v}_\Face}}
\newcommand{\Density}{\ensuremath{\rho}}
\newcommand{\Dynviscosity}{\ensuremath{\mu}}
\newcommand{\Kinviscosity}{\ensuremath{\nu}}
\newcommand{\Surfacetensioncoeff}{\ensuremath{\sigma}}
\newcommand{\Velocity}{\ensuremath{\mathbf{v}}}
\newcommand{\Pressure}{\ensuremath{p}}
\newcommand{\Res}{\ensuremath{\mathbf{r}}}
\newcommand{\Norm}[1]{\ensuremath{\left|#1\right|}}
\newcommand{\Linf}[1]{\ensuremath{L_\infty\left( #1 \right)}}
\newcommand{\Lone}[1]{\ensuremath{L_1\left( #1 \right)}}
\newcommand{\Ltwo}[1]{\ensuremath{L_2\left( #1 \right)}}
\newcommand{\Tollinsolver}{\ensuremath{\text{tol}_\text{ls}}}
\newcommand{\Tolrel}{\ensuremath{\text{tol}_\text{rel}}}
\newcommand{\Tolabs}{\ensuremath{\text{tol}_\text{abs}}}
\newcommand{\Hop}{\mathbf{H}}
\newcommand{\Dop}{\mathbf{D}}
\newcommand{\source}{\mathbf{b}}
\newcommand{\Source}{\mathbf{B}}
\newcommand{\kiter}{{(k)}}
\newcommand{\Grad}[1]{\nabla#1}
\newcommand{\dgMarkerfield}{DG($\alpha$)}
\newcommand{\dgSigneddistance}{DG($\phi$)}
\newcommand{\compactdgnc}{cDG($\phi$)}
\newcommand{\compactsphere}{sccDG($\phi$)}
\newcommand{\Saample}{SAAMPLE}
\newcommand{\ODfreq}{\omega}
\newcommand{\Decayf}{\gamma}
\newcommand{\Semix}{\ensuremath{s_x}}
\renewcommand{\arraystretch}{1.1}


\begin{frontmatter}


\title{SAAMPLE: A Segregated Accuracy-driven Algorithm for Multiphase Pressure-Linked Equations}
\author{Tobias Tolle}
\ead{tolle@mma.tu-darmstadt.de}
\author{Dieter Bothe}
\ead{bothe@mma.tu-darmstadt.de}
\author{Tomislav Mari\'{c}\corref{corr}}
\cortext[corr]{Corresponding author}
\ead{maric@mma.tu-darmstadt.de}
\address{Mathematical Modeling and Analysis, Technische \mbox{Universit{\"a}t} Darmstadt}

\begin{abstract}

\emph{Note}: this is an updated preprint of the manuscript accepted for publication in \emph{Computers \& Fluids}, DOI: \url{https://doi.org/10.1016/j.compfluid.2020.104450}. Please refer to the journal version when citing this work.

Existing hybrid Level Set / Front Tracking methods have been developed for structured meshes and successfully used for efficient and accurate simulations of complex multiphase flows. This contribution extends the capability of hybrid Level Set / Front Tracking methods towards handling surface tension driven multiphase flows using unstructured meshes. Unstructured meshes are traditionally used in Computational Fluid Dynamics to handle geometrically complex problems. In order to simulate surface-tension driven multiphase flows on unstructured meshes, a new SAAMPLE Segregated Accuracy-driven Algorithm for Multiphase Pressure-Linked Equations is proposed, that increases the robustness of the unstructured Level Set / Front Tracking (LENT) method. The LENT method is implemented in the OpenFOAM open source code for Computational Fluid Dynamics. 
\end{abstract}

\begin{keyword}
  Level Set, Front Tracking, unstructured, pressure-velocity coupling, surface tension, OpenFOAM
\end{keyword}

\end{frontmatter}



\section{Introduction}
\label{section:introduction}
\input{sections/introduction}

\section{Two-phase flow model}
\label{section:mathmodel}
\input{sections/mathematical-model}

?both\section{Numerical method}
\label{section:numerical-method}
\input{sections/numerical-method}

\section{Numerical results}
\label{section:numerical-results}
\input{sections/numerical-results}
\section{Conclusions}
\input{sections/conclusions}

\section{Acknowledgements}
\input{sections/acknowledgements}

\bibliographystyle{unsrtnat}
\bibliography{bibliography}






\end{document}

%% file: sections/introduction.tex
Multiphase flow simulations are becoming an increasingly important tool for designing and optimizing natural and technical processes. Combustion byproduct reduction in exhaust systems, ship resistance minimization, icing of airplane wings and blades of wind-power generators, fuel cell design - to name only a few important technical systems that are simulated and optimized using multiphase flow simulations. 

At their core, numerical methods for multiphase flow simulations attempt to accurately and efficiently approximate the evolution of interfaces that form between immiscible fluid phases. An accurate, stable and efficient motion of the fluid interface in the context of multiphase flows consists of two components: the kinematics of the interface and the solution of a multiphase Navier-Stokes system. 

In a previous publication \cite{Maric2015}, a new LENT hybrid Level Set / Front Tracking method was developed on unstructured meshes. This work extends the LENT method towards two-phase flows driven by the surface tension forces. For this purpose, the SAAMPLE Segregated Accuracy-driven Algorithm for Multiphase Pressure-Linked Equations is developed to stabilize for the single-field formulation of Navier-Stokes equations on unstructured meshes.  

Before the new solution algorithm of the LENT method is described, it should be placed in the context of other contemporary contributions. Research of multiphase simulation methods has produced a substantial amount of scientific contributions over the years. Here we place the focus only on the methods that are directly or indirectly related to the hybrid Level Set / Front Tracking method.

Widely used multiphase flow simulation methods can be categorized into: Front Tracking \cite{Unverdi1992,Glimm1998,Tryggvason2001}, Level Set \cite{Sethian1996,Sussman1998,Gibou2018} and Volume-of-Fluid (VOF) \cite{Hirt1981,Rider1998} methods. Each method has specific advantages and disadvantages with respect to the other methods. All methods are still very actively researched and a relatively recent research avenue is focused on hybrid methods. Hybrid methods are set to outperform original methods by combining their sub-algorithms, with the goal of combining strengths and avoiding weaknesses of individual methods.

A notable example is the widely used coupled Level Set and Volume-of-Fluid method (CLSVOF) \cite{Sussman2000}. CLSVOF was developed to address the disadvantage of the Volume-of-Fluid method in terms of accurate surface tension calculation and the disadvantage of the Level Set method in terms of volume conservation. A similar hybrid method between the Moment of Fluid (MoF) method \cite{Dyadechko2008} and the Level Set method has been developed using a collocated solution approach and block-structured adaptive mesh refinement (AMR) \cite{Jemison2013}.

A very promising hybrid method is the hybrid Level Set / Front Tracking method. Here, the Level Set method is used to simplify the handling of topological changes of the interface and improve the accuracy of the curvature approximation, while the Front Tracking method is employed for its widely known accuracy in tracking the interface. 

The Front Tracking method approximates the fluid interface using a set of mutually connected lines in 2D and triangles in 3D.  Coalescence and breakup change the connectivity of the Front and these operations are possibly global, because coalescence or breakup may involve interaction between arbitrary parts of the fluid interface. Global topological operations are therefore required to handle topological changes in the connectivity of the Front, and the corresponding changes in connectivity then complicate an efficient implementation. This especially concerns the efficiency of the parallel implementation of the Front Tracking method in non-periodic solution domains. More information about the Front Tracking method is available in \cite{Tryggvason2011}.

The hybrid Level Contour Reconstruction Method (LCRM) \cite{Shin2002, Shin2005, Shin2007, Shin2011, Shin2017} simplifies the topological changes of the interface while ensuring stability, accuracy and computational efficiency of the fluid interface motion. The connection between LCRM and the original Level Set method is the use of a signed distance field. The signed distance field is computed in the near vicinity of the Front and it is updated as the Front moves in space. A zero level set (\ie an iso-surface) reconstruction from this distance field automatically handles topological changes of the interface. Iso-surface algorithms do not require large cell stencils, so an efficient parallel implementation can be achieved using a straightforward domain decomposition approach. Other researchers have extended the hybrid Level Set / Front Tracking method with block-adaptive structured mesh refinement (block AMR). Block AMR is applied near the interface in order to increase accuracy and reduce errors in mass conservation \cite{Ceniceros2010}. Hybrid Level Set / Front Tracking has also been successfully developed using the Finite Element discretization, for fluid-solid interaction \cite{Basting2013} and two-phase flows \cite{Basting2014}. In this approach, the immersed Front is used as a surface onto which vertices of a 2D unstructured mesh are projected, to ensure the necessary alignment of face and interface normal vectors.

All the aforementioned Front Tracking and hybrid Level Set / Front Tracking methods are developed on structured meshes. Structured methods can employ very accurate interpolations and still maintain high computational efficiency \cite{Shin2017}. On structured meshes,  geometrically complex solution domains are often handled using the Immersed Boundary Method (IBM) \cite{Mittal2005}.

Unstructured meshes greatly simplify simulations of multiphase flows in geometrically complex domains, in terms of a relatively straightforward domain discretization. However, unstructured meshes also introduce additional challenges when used with hybrid Level Set / Front Tracking methods in the context of the Finite Volume method (FVM). To address the specific challenge of an accurate and stable solution of the two-phase Navier-Stokes system for the LENT method, we propose the new \Saample{} segregated solution algorithm, outlined in the following sections.

%% file: sections/mathematical-model.tex
\begin{figure}[h] 
    \centering
    \def\svgwidth{0.6\textwidth}
       {\footnotesize
        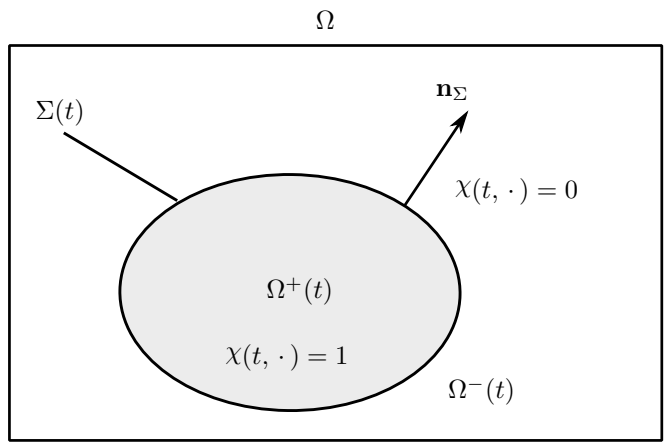
       }
       \caption{Two-phase flow solution domain.}
	\label{fig:mathmodel}
\end{figure}

Two-phase flow is modeled by a solution domain $\Omega$ filled with two immiscible incompressible phases $\Omega^+(t)$ and $\Omega^-(t)$, that are separated by a sharp interface $\Sigma(t)$ with the normal vector $\n_\Sigma$, as shown in \cref{fig:mathmodel}. The incompressible Navier-Stokes equations in a single-field formulation are used to model the flow of the two phases. The model consists of the volume (mass) conservation equation
\begin{equation}
  \div{\U} = 0, 
  \label{eq:continuity}
\end{equation}
and the momentum conservation equation in a conservative form, i.e.\ 
\begin{equation}
  \partial_t(\rho\U) + \div{\left(\rho\U\otimes\U - \mathbf{T}\right)}
       =  \rho \mathbf{g} + \Fsigma.
    \label{eq:momentum}
\end{equation}
The stress tensor for an incompressible Newtonian fluid is given as
\begin{equation}
\mathbf{T} = -\left(p + \dfrac{2}{3}\mu\div{\U}\right)\mathbf{I} + \mu(\grad{\U} + (\grad{\U})^T).
\end{equation}
The phase indicator function, used to distinguish between the two phases (cf. \cref{fig:mathmodel}), is given as 
\begin{equation}
    \rchi(t,\x) =     
    \begin{cases}    
        1 \quad \x \in \Omega^+(t), \\    
        0 \quad \x \in \Omega^-(t).    
    \end{cases}    
   \label{heavisideFunction}
\end{equation} 
The phase indicator function is used to model the single-field density and consequently the dynamic viscosity according to 
\begin{align}
   \rho = \rchi \rho_1 + (1-\rchi)\rho_2,\nonumber \\
   \Dynviscosity = \rchi \Dynviscosity_1 + (1-\rchi)\Dynviscosity_2,
   \label{eq:mixturemodel}
\end{align}
where $\rho_1$, $\rho_2$ and $\Dynviscosity_1$, $\Dynviscosity_2$ are the constant densities and dynamic viscosities of the first and the second phase, respectively. The surface tension force density \Fsigma{} is given as a volumetric source term, i.e. 
\begin{equation}
  \Fsigma = \sigma \kappa \Nsigma{} \delta_{\Sigma},
\end{equation}
where surface tension coefficient $\sigma$ is assumed constant, $\kappa$ is the interface curvature and \Nsigma{} is the unit normal to the interface $\Sigma$.

%% file: figures/mathmodel-domain.pdf_tex
\begingroup%
  \makeatletter%
  \providecommand\color[2][]{%
    \errmessage{(Inkscape) Color is used for the text in Inkscape, but the package 'color.sty' is not loaded}%
    \renewcommand\color[2][]{}%
  }%
  \providecommand\transparent[1]{%
    \errmessage{(Inkscape) Transparency is used (non-zero) for the text in Inkscape, but the package 'transparent.sty' is not loaded}%
    \renewcommand\transparent[1]{}%
  }%
  \providecommand\rotatebox[2]{#2}%
  \newcommand*\fsize{\dimexpr\f@size pt\relax}%
  \newcommand*\lineheight[1]{\fontsize{\fsize}{#1\fsize}\selectfont}%
  \ifx\svgwidth\undefined%
    \setlength{\unitlength}{189bp}%
    \ifx\svgscale\undefined%
      \relax%
    \else%
      \setlength{\unitlength}{\unitlength * \real{\svgscale}}%
    \fi%
  \else%
    \setlength{\unitlength}{\svgwidth}%
  \fi%
  \global\let\svgwidth\undefined%
  \global\let\svgscale\undefined%
  \makeatother%
  \begin{picture}(1,0.68564763)%
    \lineheight{1}%
    \setlength\tabcolsep{0pt}%
    \put(0,0){\includegraphics[width=\unitlength,page=1]{mathmodel-domain.pdf}}%
    \put(0.05992822,0.49357669){\color[rgb]{0,0,0}\makebox(0,0)[lt]{\lineheight{0}\smash{\begin{tabular}[t]{l}$\Sigma(t)$\end{tabular}}}}%
    \put(0,0){\includegraphics[width=\unitlength,page=2]{mathmodel-domain.pdf}}%
    \put(0.64673413,0.52687048){\color[rgb]{0,0,0}\makebox(0,0)[lt]{\lineheight{0}\smash{\begin{tabular}[t]{l}$\n_\Sigma$\end{tabular}}}}%
    \put(0.39857885,0.23024357){\color[rgb]{0,0,0}\makebox(0,0)[lt]{\lineheight{0}\smash{\begin{tabular}[t]{l}$\Omega^+(t)$\end{tabular}}}}%
    \put(0.66511731,0.08341812){\color[rgb]{0,0,0}\makebox(0,0)[lt]{\lineheight{0}\smash{\begin{tabular}[t]{l}$\Omega^-(t)$\end{tabular}}}}%
    \put(0,0){\includegraphics[width=\unitlength,page=3]{mathmodel-domain.pdf}}%
    \put(0.47098518,0.6249519){\color[rgb]{0,0,0}\makebox(0,0)[lt]{\lineheight{0}\smash{\begin{tabular}[t]{l}$\Omega$\end{tabular}}}}%
    \put(0.33852038,0.13395806){\color[rgb]{0,0,0}\makebox(0,0)[lt]{\lineheight{0}\smash{\begin{tabular}[t]{l}$\rchi(t,\cdot) = 1$\end{tabular}}}}%
    \put(0.67513836,0.37815904){\color[rgb]{0,0,0}\makebox(0,0)[lt]{\lineheight{0}\smash{\begin{tabular}[t]{l}$\rchi(t,\cdot) = 0$\end{tabular}}}}%
  \end{picture}%
\endgroup%

%% file: sections/numerical-method.tex
The LENT hybrid Level Set / Front Tracking method \cite{Maric2015} is used for the evolution of the interface, and the unstructured Finite Volume Method in the OpenFOAM computational fluid dynamics platform \citep{JasakPhD,JureticPhD,Moukalled2016} is used for the discretization of two-phase Navier-Stokes \cref{eq:continuity,eq:momentum}. This contribution improves the LENT method \cite{Maric2015}, in terms of the phase indicator and curvature  approximation as well as the pressure-velocity coupling algorithm. 

Algorithms of the LENT method and their respective improvements are outlined in the following sections.

\begin{figure}[h] 
    \centering
    \def\svgwidth{0.6\textwidth}
       {\footnotesize
        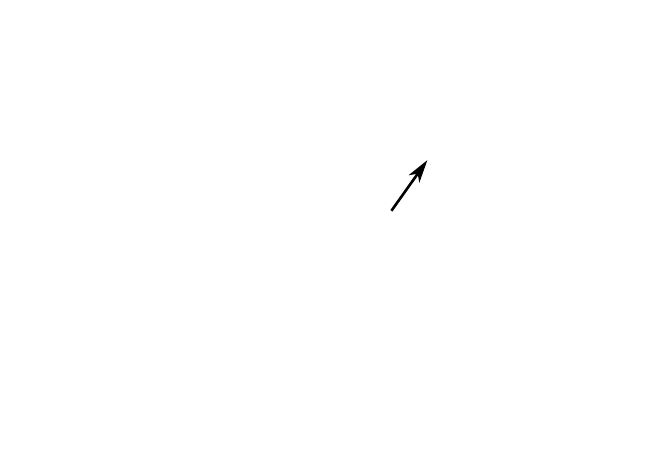
       }
       \caption{Two-phase flow domain discretization.}
	\label{fig:discrdomain}
\end{figure}

\subsection{Interface evolution}
\label{subsec:evolution}

The solution domain $\Omega$ is discretized into the discrete domain $\Omega_h$ that consists of non-overlapping polyhedral finite volumes $V_c$ (cf. \cref{fig:discrdomain}) such that $\Omega_h = \cup_{c} V_c, \, c \in C$. The LENT method further approximates the fluid interface $\Sigma(t)$ (cf.\  \cref{fig:mathmodel}) with a set of triangles $\Gamma(t)$: the so-called \emph{Front}. The motion of the Front is then given by a kinematic equation for each Front vertex $\x^k_\Gamma$, \ie 
\begin{equation}
\partial_t \x^k_\Gamma = \U(t, \x^k_\Gamma), \quad k \in K.
\label{eq:kinematic}
\end{equation}
The velocity $\U(t, \cdot)$ is obtained  from the numerical solution of \cref{eq:continuity,eq:momentum}. The Front vertex $\x^k_\Gamma$ does not in general coincide with mesh points used in the domain discretization. Therefore, the velocity $\U(t,\x^k_\Gamma)$ must be interpolated from the discretized domain (\ie unstructured mesh) at $\x^k_\Gamma$. To interpolate the velocity, the front vertices $\x^k_\Gamma$ must first be located with respect to mesh cells. This is achieved by a  combination of the octree space subdivision and known-vicinity search algorithms \cite{Maric2015}. For handling the topological changes of the interface, the iso-surface reconstruction by marching tetrahedra \cite{Treece1998} is used. 

The time step restriction imposed by the resolution of capillary waves renders higher-order methods unnecessary, which are usually used for the interpolation of $\U(t, \x^k_\Gamma)$, as well as the temporal integration of \cref{eq:kinematic}. We have therefore used Inverse Distance Weighted approximation for the velocity $\U(t,\x^k_\Gamma)$ and the explicit Euler method for the integration of \cref{eq:kinematic}, same as in \citep{Maric2015}.

\subsection{Phase indicator approximation}
\label{subsec:phase-indicator}
For the discretization of \cref{eq:momentum}, a volume averaged phase indicator
%
\begin{equation}
    \PhaseIndicator_\Cell (t) = \frac{1}{V_\Cell}
        \int_C \rchi(t,\x) \text{d}V
    \label{eq:discrete-phase-indicator}
\end{equation}
is required, e.g. to compute the material properties of cells $\Cell : \Cell \cap \Gamma \neq \emptyset$.
In the previous publication \citep{Maric2015} $\PhaseIndicator_\Cell$ is approximated with a harmonic function of $\SignedDistance$ which was also used in the LEFT hybrid level set / front tracking method \citep{Ceniceros2010}. 
The width of the marker field computed by this approach can be as large as the narrow band (4-5 cells) or limited to the single layer of cells that are intersected by the front.
Here, a method proposed in \cite{Detrixhe2015} is adopted which approximates the volume fraction in a cell from signed distances stored at the cell center and corner points. It yields a second-order accurate approximation of the volume fraction with one mixed cell ($0 < \PhaseIndicator_c < 1$) in the direction of the interface normal. In contrast, the harmonic model does not converge on a cell level.

The marker field model in \cite{Detrixhe2015} approximates the volume fraction of a tetrahedron $\Tetrahedron$ as
\begin{equation*}
    \PhaseIndicator(\Tetrahedron) = f(\SignedDistance_{e,i}), \quad i=(1,2,3,4)
\end{equation*}
where $\SignedDistance_{e,i}$ are the vertices' signed distances to the interface. So, for an arbitrary polyhedral cell $\Cell$, its phase indicator value is computed as
\begin{equation}
    \PhaseIndicator_\Cell = \frac{1}{V_c} \sum_e V(\Tetrahedron)\PhaseIndicator(\Tetrahedron)
\end{equation}
where $\Tetrahedron$ are obtained from a tetrahedral decomposition of $\Cell$. We choose the centroid decomposition used in \citep{Bloomenthal1994} as it only relies on the cell center and the cell vertices, because the signed distance is already available at these locations.

\subsubsection{Approximation of area fractions for cell-faces}
\label{subsec:wetted-area}
The discretization of the convective term in \cref{eq:momentum} requires the mass flux $\Massflux$ at cell faces $\Face$. In the LENT method $\Massflux$ needs to be computed from the volumetric flux $\Volflux$ and the density at the face $\Density_\Face$. 
While $\Density_\Face$ is simply the corresponding fluid's density for faces of bulk cells, attention must be paid for faces of interface cells.
Thus, $\Density_\Face$ is calculated analogously to the density at cell centers \cref{eq:mixturemodel} by taking an area weighted average of the bulk densities
\begin{equation}
    \Density_\Face = \PhaseIndicator_\Face \Density_1 + (1-\PhaseIndicator_\Face) \Density_2
    \label{eq:density-at-face}
\end{equation}
where $\PhaseIndicator_\Face$ denotes the fraction of face $\Face$ wetted by the corresponding phase. The area fractions are computed in a similar fashion as the volume fractions by using the two-dimensional variant of \citep{Detrixhe2015}. For the sake of efficiency, $\PhaseIndicator_\Face$ is only computed in this way for faces intersected by the Front, i.e. 
$\Face \in \Cell : \Cell \cap \Front \neq \emptyset$.

\subsection{Curvature approximation}
\label{subsec:curvatureapp}
The mean curvature $\MeanCurvature$ of the interface $\Interface$ is given by
\begin{equation}
    \MeanCurvature = -\nabla_\Interface \cdot \Nsigma{}.
    \label{eq:curvature-interface}
\end{equation}
With a level set field $\psi(\x)$ representing $\Interface$ at the iso-contour $\psi(\x) = \psi_\Interface$, \cref{eq:curvature-interface}
can be replaced by
\begin{equation}
    \MeanCurvature = -\div \frac{\grad\psi(\x)}{\Norm{\grad\psi(\x)}} \quad \text{for }
        \x \in \Interface.
    \label{eq:curvature-level-set}
\end{equation}
%
In the context of the LENT method $\psi$ is either the phase indicator $\PhaseIndicator$ or the signed distance $\SignedDistance$ field. For the sake of simplicity $\psi = \PhaseIndicator$ has been chosen in \citep{Maric2015} for a preliminary coupling of LENT with the Navier-Stokes equations. The present work, however, uses $\SignedDistance$ for the calculation of $\MeanCurvature$.

\begin{figure}
    \centering
    \begin{tikzpicture}
    \begin{axis} [
            xlabel = {$\SignedDistance$},
            ylabel = {$\MeanCurvature$},
            grid = major
        ]
        \addplot [domain=-0.2:0.2] {2/(1 + x)};
    \end{axis}
    \end{tikzpicture}
    \caption{Curvature of a sphere with $R=1$, evaluated in the normal direction to the sphere, as a function of the signed distance from the sphere $\SignedDistance$ using \cref{eq:curvature-level-set}. Obviously, the exact value $\MeanCurvature = 2$ can only be obtained on the sphere, where $\SignedDistance = 0$.}
    \label{fig:unit-sphere-level-set-curvature}
\end{figure}
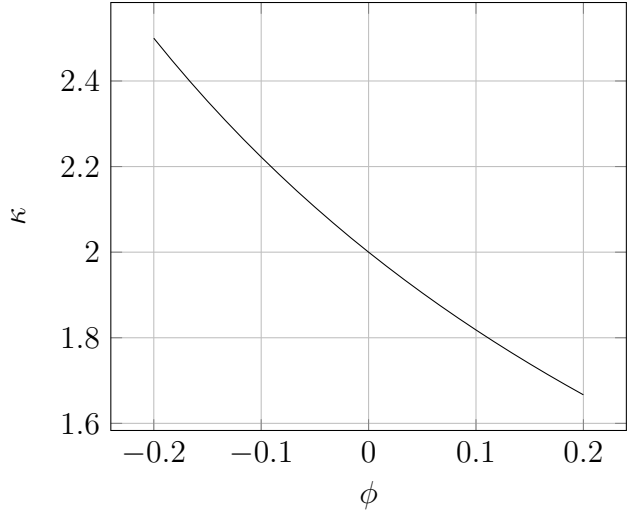

As pointed out in \citep{Sussman2009} and \citep{Popinet2017}, \cref{eq:curvature-level-set} does not yield the curvature of the interface if $\x \notin \Interface$.
Instead, \cref{eq:curvature-level-set} gives the curvature of the contour that passes through the point where \cref{eq:curvature-level-set} is evaluated. For example, at a cell center, the curvature of a contour $\SignedDistance = \SignedDistance_C$ is computed. Thus, $\MeanCurvature$ changes in normal direction of the interface as illustrated for a sphere in \cref{fig:unit-sphere-level-set-curvature}.
This error can be mitigated to some degree by the so-called \emph{compact curvature calculation}. 

\subsubsection{Comapct curvature calculation}

\begin{figure}[h] 
    \centering
    \def\svgwidth{0.6\textwidth}
       {\footnotesize
        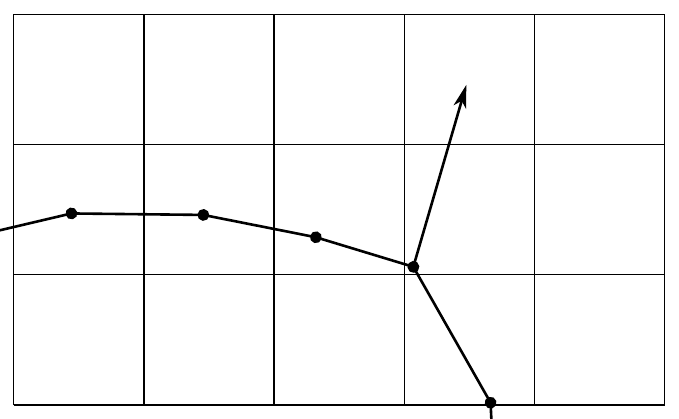
       }
       \caption{Compact curvature calculation.}
	\label{fig:compactkappa}
\end{figure}

\label{subsec:compactcurvature}
In \citep{ShinJuric2009} the authors introduce the concept of \emph{compact curvature calculation} for their LCRM method to propagate the curvature in the narrow band as follows.  A different approach to the actual curvature approximation is used in \citep{ShinJuric2009}, so we adopt the compact curvature calculation only for the correction of the curvature in the narrow band of cells surrounding the Front. 

Consider the example Front and narrow band configuration in \cref{fig:compactkappa}. Each cell center $\Point_c$ in the narrow band is associated with a point $\Point_m \in \Front$ such that the distance between $\Point_c$ and 
$\Front$ is minimal. Using this connection $\MeanCurvature$ is first interpolated to $\Point_m$, and then $\MeanCurvature(\Point_c) = \MeanCurvature(\Point_m)$ is set.

In the LENT method, $\MeanCurvature$ is first computed using \cref{eq:curvature-level-set}. This gives a curvature field that varies along interface normal direction in general (cf. \cref{fig:unit-sphere-level-set-curvature}). To reduce the curvature variation in the interface normal direction, only the curvature values computed at cells $\Cell$ for which $\Front \cap \Cell \neq \emptyset$ are kept. This information is propagated in approximate interface normal direction by combining two maps. First
\begin{equation}
    \TriangleMap(k) : k \rightarrow \Triangle_m 
    \label{eq:map_cell_to_triangle}
\end{equation}
gives the closest triangle $\Triangle_m$ for each cell $k$ in the narrow band \emph{not intersected by the Front}, i.e.\ $\Front \cap k = \emptyset$. The second map
\begin{equation}
    \mathcal{M}_c(m) :\Triangle_m \rightarrow c 
    \label{eq:map_triangle_to_cell}
\end{equation}
associates the triangle $\Triangle_m$ with a narrow-band cell $\Cell$, \emph{that is intersected by the Front} and is nearest to $\Triangle_m$. Taken together, the maps relate a non-intersected narrow band cell $k$ to an interface cell $\Cell$, the nearest Front-intersected narrow band cell. Now the curvature of the nearest non-intersected cell $k$ is set as
\begin{equation}
    \MeanCurvature_k = \MeanCurvature_\Cell.
    \label{eq:compact_curvature}
\end{equation}
Results presented in \cref{subsection:curvature-results} confirm that this compact calculation improves the curvature approximation considerably in terms of accuracy. Yet, the source of error does not vanish since $\SignedDistance(\Cell) \neq 0$ in general. The maximum error can be estimated for a spherical interface of radius $R_\Sigma$ and a cubic cell with an edge length $h$. The radius of the bounding sphere of this cell is
$R_\text{bs} = \sqrt{3}h/2$. This is also the maximum distance between the interface and an interface cell since
\begin{equation}
    \Norm{\SignedDistance(\Cell)} \leq R_\text{bs}\quad
    \forall\quad
    \Cell : \Front \cap \Cell \neq \emptyset. 
\end{equation}
The exact curvature of a sphere is given by
\begin{equation}
    \MeanCurvature = \frac{2}{R},
    \label{eq:curvature_sphere}
\end{equation}
while the approximate curvature is
\begin{equation}
    \widetilde{\MeanCurvature} = \frac{2}{R_\Sigma + \SignedDistance}.
    \label{eq:approximate_curvature_sphere}
\end{equation}
Thus, the relative curvature error is given by
\begin{equation}
    e_{\MeanCurvature,\text{rel}}(\SignedDistance) =
    \frac{\Norm{\widetilde{\MeanCurvature} - \MeanCurvature}}{\MeanCurvature} =
    \left| \frac{-\SignedDistance}{R_\Sigma + \SignedDistance}\right|.
    \label{eq:relative_curvature_error_sphere}
\end{equation}
Setting $\SignedDistance = -R_\text{bs} = -\sqrt{3}h/2$ and expressing
$h=R_\Sigma/n$, where $n$ is the number of cells per radius, yields
\begin{equation}
    e_{\MeanCurvature,\text{rel}}(n) = \frac{\sqrt{3}}{2n - \sqrt{3}}.
    \label{eq:relative_curvature_error_n}
\end{equation}
This indicates first order convergence of the maximum relative curvature error with respect to mesh resolution. To further reduce the curvatue error we employ an additional correction for the curvature computed at interface cells.

\subsubsection{Spherical curvature correction}
\label{subsec:spherical_curvature_correction}
Though the interface will not be spherical in the general case, we propose a correction assuming the interface to be locally spherical due to the following observations:
%
%
\begin{enumerate}[(i)]
    \item The proposed correction is consistent and vanishes in the limiting
        case $\SignedDistance_\Cell \rightarrow 0$.
    \item Accurate curvature approximation becomes more important with more dominant surface tension often involving close to spherical interface configurations.
\end{enumerate}
If the interface is assumed to be locally spherical, the curvature error introduced by the cell signed distance $\SignedDistance_\Cell$ can be remedied in a rather simple way. If the initial curvature $\widetilde{\MeanCurvature_\Cell}$ is given by the compact curvature correction, \cref{eq:approximate_curvature_sphere} can be used to compute an equivalent interface radius $R_\Sigma$. Inserting $R_\Sigma$ in \cref{eq:curvature_sphere} yields a distance-corrected curvature
\begin{equation}
    \MeanCurvature_\Cell = 2\left( \frac{2}{\widetilde{\MeanCurvature_\Cell}} + \SignedDistance_\Cell \right)^{-1}.
    \label{eq:spherical_correction}
\end{equation}

\subsection{Surface tension force reconstruction}
\label{subsec:num:forcerecon}
We use a semi-implicit surface tension model, proposed by Raessi et al. \citep{Raessi2009} which is an extension of the original Continuum Surface Force (CSF) model of Brackbill et al. \citep{Brackbill1992}. In \citep{Raessi2009}, the surface tension is modeled as
\begin{equation}
    \Fsigma^{n+1} = \Surfacetensioncoeff (\MeanCurvature \Nsigma{})^n \delta_{\Sigma}
        + \Surfacetensioncoeff\Delta t (\LBop\Velocity{^{n+1}})\delta_{\Sigma}
    \label{eq:semi-implicit-csf}
\end{equation}
when a backward Euler scheme is used for temporal discretization. $\LBop$ denotes the Laplace-Beltrami operator. As in the original CSF we employ the approximations $\delta_{\Sigma} \approx \Norm{\Grad{\PhaseIndicator}}$ and $\Nsigma{}\delta_{\Sigma} \approx \Grad{\PhaseIndicator}$.
To achieve an implicit discretization of the Laplace-Beltrami operator we use the representation
\begin{equation}
    \LBop \Velocity = \underline{\laplacian{\Velocity}} - \div{\left[ (\Nsigma{} \cdot \Grad{\Velocity})\otimes\Nsigma{}\right]} - \MeanCurvature\left[ \left( \Grad{\Velocity} - (\Nsigma{} \cdot \Grad{\Velocity})\otimes \Nsigma{})\right)\cdot \Nsigma{}\right].
    \label{eq:laplace-beltrami-decomposition}
\end{equation}
The underlined term $\underline{\laplacian{\Velocity}}$ is discretized implicitly while the remaining terms are treated explicitly. Since an iterative approach is used to solve the pressure velocity system (see \cref{subsec:pucoupling}), the converged solution is not affected. 

However, the explicit contribution to the surface tension force requires additional attention. The pseudo-staggered unstructured FVM stores scalar numerical flux values at face centers. However, the solution algorithm requires cell-centered vector values for the surface tension force in the momentum equation. Therefore, the explicit surface tension force term is reconstructed from face-centered scalar flux values. The reconstruction operator that approximates $\phib_c$ in OpenFOAM is given as
\begin{equation}
  \phib^R_c \approx \Reconstruct(\phib_f) = \left[ \sum_{\tilde{f}} \mathbf{S}_{\tilde{f}} \mathbf{S}_{\tilde{f}}  \right]^{-1} \cdot \sum_f \Sfhat \cdot \Sf \cdot \phib_f = S_c^{-1} \cdot  \sum_f \Sfhat \cdot \Sf \cdot \phib_f,
  \label{eq:reconstruct}
\end{equation}
where $\phib^R_c$ is the cell-centered reconstructed vector value, $f$ is the index of a polygonal face that belongs to the polyhedral cell $c$, 
%
%
and $\phib^R_c$ is the reconstructed vector value, associated with the centroid of the polyhedral cell $c$ and $\Sfhat,\Sf$ are the respective outward oriented unit normal and normal vector of face $f$. Usually, the vector quantity $\phib_f$ is not available, otherwise it would be possible to compute $\phib_c$ using interpolation. Instead, the product $\phib_f \cdot \Sf$ is given, namely the \emph{scalar flux} of the vector quantity $\phib$ through the face $f$. The reconstruction operator introduces an error $\mathbf{\epsilon}_R$ in \cref{eq:reconstruct} as
\begin{equation}
    \phib_c =  S_c^{-1} \cdot \sum_f \Sfhat \cdot  \Sf \cdot  \phib_f+ \mathbf{\epsilon}_R.
    \label{eq:reconstructerr}
\end{equation}
The reconstruction error is thus expressed as 
\begin{equation}
    \mathbf{\epsilon}_c^R =  S_c^{-1} \cdot \sum_f \Sfhat \cdot \Sf \cdot (\phib_c - \phib_f).
  \label{eq:recerrorex}
\end{equation}
\Cref{eq:recerrorex} is the exact equation for the error introduced by the reconstruction operator given by \cref{eq:reconstruct}. If $\phib_f$ is sufficiently smooth, we can write 
\begin{equation}
  \phib_f = \phib_c + \grad\phib|_c\cdot\cf + \nabla\nabla\phib|_c:(\cf \cf) + \dots,
  \label{eq:taylorphi}
\end{equation}
and \cref{eq:taylorphi} can be used to replace $\phib_c - \phib_f$  in \cref{eq:recerrorex}. 

\begin{figure}[h] 
    \centering
    \def\svgwidth{0.4\textwidth}
       {\footnotesize
        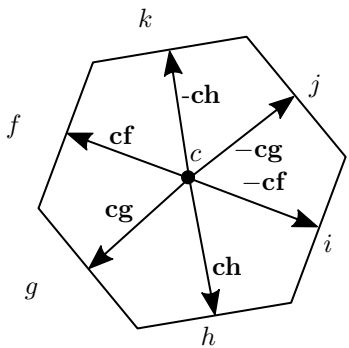
       }
       \caption{The cell-face connectivity of a regular hexagonal cell.}
	\label{fig:cf-connection}
\end{figure}

The vector $\mathbf{cf} = \mathbf{x_f} - \mathbf{x}_c$ shown in \cref{fig:cf-connection} connects the centroid of the cell $c$ and the centroid of the face $f$. Inserting \cref{eq:taylorphi} into \cref{eq:recerrorex}, while disregarding higher-order terms, leads to
\begin{align}
    \tilde{\mathbf{\epsilon}}_c^R & = -S_c^{-1} \cdot (\sum_f \Sfhat \cdot  \Sf \cdot \grad\phib|_c\cdot \cf + \sum_f \Sfhat \cdot  \Sf \cdot \nabla\nabla\phib|_c : (\cf\cf)).
  \label{eqn:recapprox}
\end{align}
The first sum cancels out in convex orthogonal cells with an even number of faces. This can be easily shown if one considers \cref{fig:cf-connection}. Regular polygons and polyhedra exist, that have an even number of faces and form so-called \emph{orthogonal unstructured meshes}\footnote{A mesh is orthogonal if the face normal vectors are  collinear with the line segments between face-neighboring cell centroids.}. For each face $i$ of such cells, there exists another face $j$, such that $\mathbf{ci} = -\mathbf{cj}$ and $\mathbf{S}_i = -\mathbf{S}_j$. Because these cells have an even number of faces, the first sum in \cref{eqn:recapprox} can be split into two sums of equal length. This, together with the opposite sign of the $\mathbf{S}_j$ and $\mathbf{cj}$ vectors leads to 
\begin{align}
    \sum_f \Sfhat \cdot  \Sf \cdot \grad\phib|_c\cdot \cf & =
            \sum_i \mathbf{\hat{S}}_i \cdot \mathbf{S}_i \cdot \grad\phib|_c\cdot \mathbf{ci} +  
            \sum_j \mathbf{\hat{S}}_j \cdot \mathbf{S}_j \cdot \grad\phib|_c\cdot \mathbf{cj}   
     \\
        & = 
             \sum_i ( \mathbf{\hat{S}}_i \cdot \mathbf{S}_i \cdot \grad\phib|_c\cdot
             \mathbf{ci} 
             -\mathbf{\hat{S}}_i \cdot -\mathbf{S}_i \cdot \grad\phib|_c\cdot
             -\mathbf{ci})
        = \mathbf{0}.
  \label{eqn:recapproxcancel}
\end{align}

The cancellation in \cref{eqn:recapproxcancel} happens for quadratic and hexagonal cells in $2D$ and hexahedral cells with orthogonal faces, as well as regular dodecahedron cells. For slightly irregular cells (e.g. hexahedrons with slightly non-orthogonal faces), the error cancellation is partial, but it may still be strong, due to the fact that $\mathbf{ct}$ is almost, but not quite, equal to $-\mathbf{cf}$. The canceling rate thus directly depends on the mesh non-orthogonality.

Therefore, for orthogonal meshes, the reconstruction error is given as
\begin{align}
    \tilde{\mathbf{\epsilon}}_c^{R,orth} & = -S_c^{-1} \sum_f \Sfhat \cdot  \Sf \cdot \nabla\nabla\phib|_c : (\cf\cf).
  \label{eqn:recorth}
\end{align}
\Cref{eqn:recorth} shows that the reconstruction of the surface tension force is second-order accurate on orthogonal unstructured meshes: the reconstruction error $\mathbf{\tilde{\epsilon}}_c^{R,orth}$ is zero if the vector field is linear. On non-orthogonal meshes, the $\cf$ vectors do not cancel out and the reconstruction may deteriorate to first order of accuracy, given by \cref{eq:recerrorex}, depending on mesh non-orthogonality.

\subsection{The SAAMPLE segregated solution algorithm}
\label{subsec:pucoupling}
In order to solve the discretized pressure-velocity system a new segregated solution algorithm based on the PISO approach \citep{Issa1986} is developed. 
An overview of the different pressure-correction methods and how they are related can be found in \citep{Darwish2001}. Barton \citep{Barton1998} compares several PISO and SIMPLE based solution procedures in terms of accuracy, robustness and computational efficiency. He concludes that PISO is the preferred algorithm for transient flows considering all metrics. This is the reasoning behind using the PISO algorithm in \citep{Maric2015}, however with $n=4$ pressure correction iterations and followed by an additional solution of the momentum equation with the updated pressure.

Given that PISO has originally been proposed as a solution procedure for single-phase flows, some drawbacks become apparent when it is used for two-phase flows. There is no control over the solution accuracy because the PISO algorithm is controlled by a fixed iteration count. How this may manifest, is demonstrated in \cref{subsection:stationary-droplet}. In the OpenFOAM framework, used to develop the LENT method, the explicit velocity update involves the use of the reconstruction operator given by \cref{eq:reconstruct}, that introduces additional errors given by \cref{eqn:recorth}. Another problem that is emphasized by multiphase flows is the inability of the PISO algorithm to account for the non-linearity of the convective term. The main contribution of this work is the alleviation of these issues in the context of surface-tension driven multiphase flows. 

Here, we propose a new PISO-based solution algorithm termed
\Saample{} (Segregated Accuracy-driven Approach for Multiphase Pressure Linked Equations) that overcomes these disadvantages. Additionally, SAAMPLE avoids the use of case-dependent parameters like under-relaxation factors of the original SIMPLE method \citep{Patankar1972}. \Saample{} is outlined in \cref{alg:pu-coupling}.
\begin{algorithm}
    \centering
    \caption{Pseudo code of the \Saample{} algorithm for the segregated solution of the pressure-velocity system. The operator ':=' denotes assignment.}
    \label{alg:pu-coupling}
    {\small
    \begin{algorithmic}[1]
        \State conv-vol-fluxes := \verb+False+
        \State pU-converged := \verb+False+
        \State $K := 0$
        \State $\Velocity := \Velocity^n$
        \State $\Pressure := \Pressure^n$
        \State \quad
        \While{\textbf{not} pU-converged \textbf{and} $K < K_\text{max}$}
            \If{\textbf{not} conv-vol-fluxes}
                \State Update mass flux: $\Massflux := \Density_f \Volflux$
                    (\cref{eq:density-at-face})  
            \EndIf
            
            \State conv-vol-fluxes := \cref{eq:flux-convergence}
            \State Solve momentum predictor   
                \cref{eq:momentum-predictor}: $\Velocity{} := \Velocity{}^*$
            
            \State \quad
            \State $I:=0$
            \State correct-pressure := \verb+True+
            \While{correct-pressure \textbf{and} $I < I_\text{max}$}
                \State Setup pressure-correction \cref{eq:pressure-correction}
                \State Compute $r$ as norm of initial residual $\Res$
                
                \If{$r > \Tollinsolver$}
                    \State Solve for $\Pressure$: $\Pressure := \Pressure^*$
                    \State Update volumetric fluxes $\Volflux$
                    \State Explicit velocity update \cref{eq:velocity-update}: $\Velocity := \Velocity^{**}$
                \Else
                    \State correct-pressure := \verb+False+
                    \If{$I=0$ \textbf{and} conv-vol-fluxes}
                        \State pU-converged := \verb+True+
                    \EndIf
                \EndIf
                \State $I := I + 1$
            \EndWhile
            \State $K:=K+1$
        \EndWhile
        \State \quad
        \State $\Velocity^{n+1} := \Velocity$
        \State $\Pressure^{n+1} := \Pressure$
    \end{algorithmic}
    }
\end{algorithm}
It employs the same equations as the orignal PISO algorithm, namely a momentum predictor (\cref{eq:momentum-predictor}), a pressure correction equation (\cref{eq:pressure-correction}) and an explicit velocity update (\cref{eq:velocity-update}), given here in a semi-discrete form as
\begin{align}
    \label{eq:momentum-predictor}
    \begin{split}
        a^\Velocity_\Cell \Velocity_\Cell + \sum_{N(\Cell)} a^\Velocity_N \Velocity_N &=  \source^\Velocity_\Cell - V_\Cell(\Grad{\Pressure})_\Cell, \\
        \Velocity^*_\Cell + \Hop_\Cell[\Velocity^*] &= - \Dop^ \Velocity_\Cell(\Grad{\Pressure}^\text{prev})_\Cell + \Source^\Velocity_\Cell,
    \end{split}\\
    \div{\left(\Dop^\Velocity_\Cell(\Grad{\Pressure^*})_\Cell\right)} &=
        \div{\left( \Hop_\Cell[\Velocity^*] + \underline{\Hop_\Cell[\Velocity']} - \Source^\Velocity_\Cell \right)},
    \label{eq:pressure-correction} \\
    \Velocity^{**} &= -\Hop_\Cell[\Velocity^*] - \Dop^ \Velocity_\Cell(\Grad{\Pressure^*})_\Cell + \Source^\Velocity_\Cell,
    \label{eq:velocity-update}
\end{align}
in whh
\begin{equation*}
    (\grad{\Pressure})_\Cell = \frac{1}{V_\Cell}\int_{V_\Cell}\Grad{\Pressure}\mathrm{d}V, \quad
    \Hop_\Cell [\Velocity] = \frac{1}{a^\Velocity_\Cell}\sum_{N(\Cell)}a^\Velocity_N \Velocity_N, \quad
    \Source^\Velocity_\Cell = \frac{\source^\Velocity_\Cell}{a^\Velocity_\Cell}, \quad
    \Dop^\Velocity_\Cell = \frac{V_\Cell}{a^\Velocity_\Cell}.
\end{equation*}
As in PISO, the underlined term $\underline{\Hop_\Cell[\Velocity']}$ in \cref{eq:pressure-correction} is neglected as the velocity corrections $\Velocity'$ are unknown.

Contrary to PISO, \Saample{} is an iterative algorithm that is driven by the solution accuracy. It consists of two nested loops with specific purposes. The outer loop updates the mass fluxes $\Massflux$ as long as the function
\begin{equation}
    \text{conv}(\Volflux) = 
        \begin{cases}
            1,\quad \text{if } \Linf{\Norm{\Volflux - \Volflux^\text{prev}}/\Linf{\Volflux}}     < \Tolrel \\
            1,\quad \text{if } \Linf{\Norm{\Volflux - \Volflux^\text{prev}}} <         \Tolabs \\
            0,\quad \text{otherwise}
        \end{cases}
    \label{eq:flux-convergence}
\end{equation}
evaluates to $0$. The parameters $\Tolrel$ and $\Tolabs$ are presribed tolerances for the relative and absolute change of $\Volflux$ between two consecutive outer iterations. Subsequently, the momentum predictor \cref{eq:momentum-predictor} is solved with a known pressure field, either from a previous time step or previous outer iteration.

The inner loop performs the pressure correction to enforce discrete volume conservation. This is achieved with a series of corrector steps as in the original PISO algorithm \citep{Issa1986}.
First, \cref{eq:pressure-correction} is solved implicitly for $\Pressure$. The Laplacian operator on the left hand side is discretized using surface normal gradients at each cell face as implemented in OpenFOAM. Subsequently, $\Velocity$ is updated explicitly according to \cref{eq:velocity-update}
As reported in \citep{Darwish2001}, this removes the need for underrelaxation as each corrector iteration partly recovers the neglected term of \cref{eq:pressure-correction}. This agrees with the findings of Venier et al. \citep{Venier2017}. They investigate the stability of the PISO algorithm using Fourier analysis and conclude that more corrector iterations provide a stronger coupling of pressure and velocity.
Iteration of the inner loop is stopped if either the initial residuals of the pressure equation are below a prescribed threshold $\Tollinsolver$ or the maximum number of inner iterations $I_\text{max}$ is exceeded.

The overall algorithm is considered converged when condition \cref{eq:flux-convergence} has been fulfilled and for the initial residual $r$ of the pressure equation in the first iteration of the inner loop $r < \Tollinsolver$ is fulfilled. It means that $\Velocity^*$ obtained from momentum predictor \cref{eq:momentum-predictor} satisfies $\div\Velocity^* < \Tollinsolver$ in a discrete sense. If convergence is not reached within $K_\text{max}$ outer iterations, the current fields for $\Velocity$ and $\Pressure$ are considered as solutions for the time step.

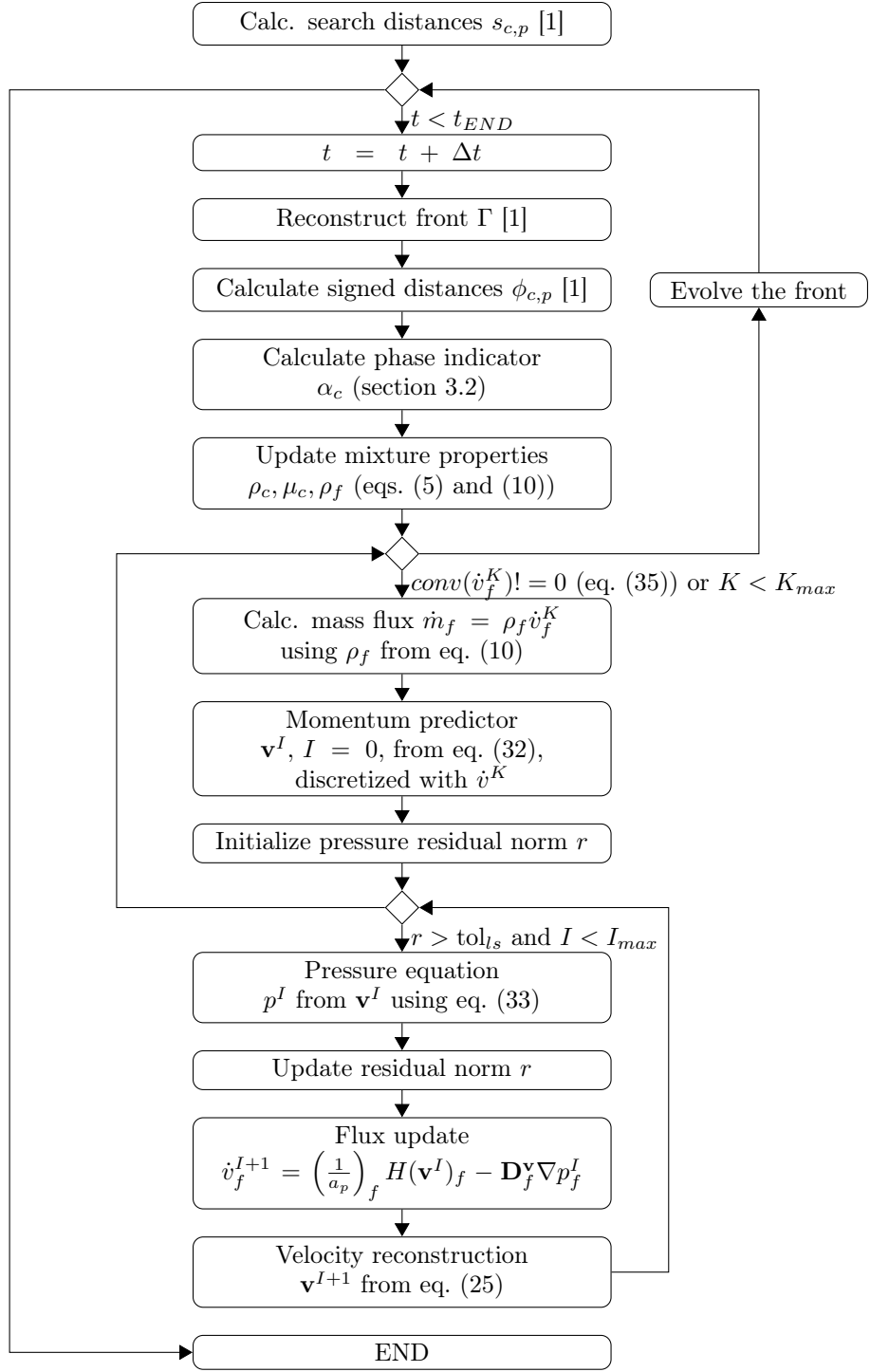
\begin{figure}[H]
    \footnotesize
    \hspace{5em}
    \begin{tikzpicture}[%
        >=triangle 60,              
        start chain=going below,    
        node distance=3.8mm and 50mm, 
        every join/.style={norm}    
        ]
        \tikzset{
          base/.style={draw, on chain, on grid, align=center},
          rounded/.style={base, rounded corners, text width=16em}, 
          test/.style={base, diamond, aspect=1}, 
          arrow/.style={->, draw}
        }
        \node [rounded] (p0) {Calc. search distances $s_{c,p}$ \citep{Maric2015}};
        \node [test]    (p1) {};
        \node [rounded] (p2) {$t = t + \Delta t$};
        \node [rounded] (p3) {Reconstruct front $\Front$ \citep{Maric2015}};
        \node [rounded] (p4) {Calculate signed distances $\phi_{c,p}$ \citep{Maric2015}};
        \node [rounded] (p5) {Calculate phase indicator\\$\alpha_{c}$ (\cref{subsec:phase-indicator})};
        \node [rounded] (p6) {Update mixture properties $\rho_{c}, \mu_c, \rho_f$ (\cref{eq:mixturemodel,eq:density-at-face})};
        \node [test]    (p7) {};
        \node [rounded] (p8) {Calc. mass flux $\dot{m}_f = \rho_f \dot{v}_f^K$ using $\rho_f$ from \cref{eq:density-at-face}};
        \node [rounded] (p9) {Momentum predictor\\ $\U^I, \, I = 0$, from \cref{eq:momentum-predictor}, discretized with $\dot{v}^K$};
        \node [rounded] (p10) {Initialize pressure residual norm $r$};
        \node [test]    (p11) {};
        \node [rounded] (p12) {Pressure equation\\ $p^I$ from $\U^I$ using \cref{eq:pressure-correction}};
        \node [rounded] (p13) {Update residual norm $r$};
        \node [rounded] (p14) {Flux update \\$\dot{v}_f^{I+1} = \left(\frac{1}{a_p}\right)_f H(\U^I)_f - \Dop^\Velocity_f\Grad{\Pressure^I}_f$};
        \node [rounded] (p15) {Velocity reconstruction\\$\U^{I+1}$ from \cref{eq:reconstructerr}};
        \node [rounded] (p16) {END};
        \node [rounded, text width=8em] (p17) [right=of p4] {Evolve the front};

        \draw[->] (p0)  --  (p1);
        \draw[->] (p1)  -- node[anchor=west] {$t < t_{END}$} (p2);
        \draw[->] (p2)  --  (p3);
        \draw[->] (p3)  --  (p4);
        \draw[->] (p4)  --  (p5);
        \draw[->] (p5)  --  (p6);
        \draw[->] (p6)  --  (p7);
        \draw[->] (p7)  -- node[anchor=west] {$conv(\dot{v}_f^K) != 0$ (\cref{eq:flux-convergence}) or $K < K_{max}$} (p8);
        \draw[->] (p8)  --  (p9);
        \draw[->] (p9)  --  (p10);
        \draw[->] (p10)  -- (p11);
        \draw[->] (p11) -- node[anchor=west] {$r > \text{tol}_{ls}$ and $I < I_{max}$} (p12);
        \draw[->] (p12) --  (p13);
        \draw[->] (p13) --  (p14);
        \draw[->] (p14) --  (p15);

        \draw[->] (p15.east) -| ++(8mm,0) |- (p11.east);
        \draw[->] (p11) -| ++(-40mm,0) |- (p7);
        \draw[->] (p7) -| (p17.south);
        \draw[->] (p17.north) |- (p1);
        \draw[->] (p1) -| ++(-55mm,0) |- (p16.west);

    \end{tikzpicture}

    \caption{Flowchart of the LENT-SAAMPLE method.}
    \label{fig:flowchart}
\end{figure}

The LENT method is outlined together with the SAAMPLE algorithm in \cref{fig:flowchart}. Algorithms of the LENT-SAAMPLE method that are not modified with respect to the previous publication \citep{Maric2015} are accordingly referenced. \Cref{fig:flowchart} shows the difference in controling the convergence between the SAAMPLE and the PISO internal loop, in terms of disregarding a fixed number of iterations and relying on the pressure residual error norm.
To prevent the decoupling of the acceleration from the forces acting at the interface, the cell-centered velocity is reconstructed using the operator defined by \cref{eq:reconstruct}.

%% file: figures/mathmodel-domain-discretization.pdf_tex
\begingroup%
  \makeatletter%
  \providecommand\color[2][]{%
    \errmessage{(Inkscape) Color is used for the text in Inkscape, but the package 'color.sty' is not loaded}%
    \renewcommand\color[2][]{}%
  }%
  \providecommand\transparent[1]{%
    \errmessage{(Inkscape) Transparency is used (non-zero) for the text in Inkscape, but the package 'transparent.sty' is not loaded}%
    \renewcommand\transparent[1]{}%
  }%
  \providecommand\rotatebox[2]{#2}%
  \newcommand*\fsize{\dimexpr\f@size pt\relax}%
  \newcommand*\lineheight[1]{\fontsize{\fsize}{#1\fsize}\selectfont}%
  \ifx\svgwidth\undefined%
    \setlength{\unitlength}{189bp}%
    \ifx\svgscale\undefined%
      \relax%
    \else%
      \setlength{\unitlength}{\unitlength * \real{\svgscale}}%
    \fi%
  \else%
    \setlength{\unitlength}{\svgwidth}%
  \fi%
  \global\let\svgwidth\undefined%
  \global\let\svgscale\undefined%
  \makeatother%
  \begin{picture}(1,0.68564763)%
    \lineheight{1}%
    \setlength\tabcolsep{0pt}%
    \put(0.10887624,0.4568497){\color[rgb]{0,0,0}\makebox(0,0)[lt]{\lineheight{0}\smash{\begin{tabular}[t]{l}$\Gamma(t)$\end{tabular}}}}%
    \put(0,0){\includegraphics[width=\unitlength,page=1]{mathmodel-domain-discretization.pdf}}%
    \put(0.62590938,0.45285954){\color[rgb]{0,0,0}\makebox(0,0)[lt]{\lineheight{0}\smash{\begin{tabular}[t]{l}$\n_\Gamma$\end{tabular}}}}%
    \put(0,0){\includegraphics[width=\unitlength,page=2]{mathmodel-domain-discretization.pdf}}%
    \put(0.47098518,0.6389818){\color[rgb]{0,0,0}\makebox(0,0)[lt]{\lineheight{0}\smash{\begin{tabular}[t]{l}$\Omega_h$\end{tabular}}}}%
    \put(0,0){\includegraphics[width=\unitlength,page=3]{mathmodel-domain-discretization.pdf}}%
    \put(0.2036156,0.21688364){\color[rgb]{0,0,0}\makebox(0,0)[lt]{\lineheight{0}\smash{\begin{tabular}[t]{l}$\x^k_\Gamma$\end{tabular}}}}%
    \put(0,0){\includegraphics[width=\unitlength,page=4]{mathmodel-domain-discretization.pdf}}%
  \end{picture}%
\endgroup%

%% file: figures/compact-curvature-calculation.pdf_tex
\begingroup%
  \makeatletter%
  \providecommand\color[2][]{%
    \errmessage{(Inkscape) Color is used for the text in Inkscape, but the package 'color.sty' is not loaded}%
    \renewcommand\color[2][]{}%
  }%
  \providecommand\transparent[1]{%
    \errmessage{(Inkscape) Transparency is used (non-zero) for the text in Inkscape, but the package 'transparent.sty' is not loaded}%
    \renewcommand\transparent[1]{}%
  }%
  \providecommand\rotatebox[2]{#2}%
  \newcommand*\fsize{\dimexpr\f@size pt\relax}%
  \newcommand*\lineheight[1]{\fontsize{\fsize}{#1\fsize}\selectfont}%
  \ifx\svgwidth\undefined%
    \setlength{\unitlength}{195.375bp}%
    \ifx\svgscale\undefined%
      \relax%
    \else%
      \setlength{\unitlength}{\unitlength * \real{\svgscale}}%
    \fi%
  \else%
    \setlength{\unitlength}{\svgwidth}%
  \fi%
  \global\let\svgwidth\undefined%
  \global\let\svgscale\undefined%
  \makeatother%
  \begin{picture}(1,0.61708248)%
    \lineheight{1}%
    \setlength\tabcolsep{0pt}%
    \put(0,0){\includegraphics[width=\unitlength,page=1]{compact-curvature-calculation.pdf}}%
    \put(0.7449545,0.55965112){\color[rgb]{0,0,0}\makebox(0,0)[lt]{\lineheight{0}\smash{\begin{tabular}[t]{l}$k$\end{tabular}}}}%
    \put(0,0){\includegraphics[width=\unitlength,page=2]{compact-curvature-calculation.pdf}}%
    \put(0.71164742,0.47084325){\color[rgb]{0,0,0}\makebox(0,0)[lt]{\lineheight{0}\smash{\begin{tabular}[t]{l}$\x_k$\end{tabular}}}}%
    \put(0.5299735,0.26168138){\color[rgb]{0,0,0}\makebox(0,0)[lt]{\lineheight{0}\smash{\begin{tabular}[t]{l}$\Triangle_m$\end{tabular}}}}%
    \put(0,0){\includegraphics[width=\unitlength,page=3]{compact-curvature-calculation.pdf}}%
    \put(0.74088287,0.36549017){\color[rgb]{0,0,0}\makebox(0,0)[lt]{\lineheight{0}\smash{\begin{tabular}[t]{l}$c$\end{tabular}}}}%
    \put(0,0){\includegraphics[width=\unitlength,page=4]{compact-curvature-calculation.pdf}}%
    \put(0.71164742,0.28046662){\color[rgb]{0,0,0}\makebox(0,0)[lt]{\lineheight{0}\smash{\begin{tabular}[t]{l}$\kappa_c$\end{tabular}}}}%
  \end{picture}%
\endgroup%

%% file: figures/cf-connection.pdf_tex
\begingroup%
  \makeatletter%
  \providecommand\color[2][]{%
    \errmessage{(Inkscape) Color is used for the text in Inkscape, but the package 'color.sty' is not loaded}%
    \renewcommand\color[2][]{}%
  }%
  \providecommand\transparent[1]{%
    \errmessage{(Inkscape) Transparency is used (non-zero) for the text in Inkscape, but the package 'transparent.sty' is not loaded}%
    \renewcommand\transparent[1]{}%
  }%
  \providecommand\rotatebox[2]{#2}%
  \newcommand*\fsize{\dimexpr\f@size pt\relax}%
  \newcommand*\lineheight[1]{\fontsize{\fsize}{#1\fsize}\selectfont}%
  \ifx\svgwidth\undefined%
    \setlength{\unitlength}{170.07874016bp}%
    \ifx\svgscale\undefined%
      \relax%
    \else%
      \setlength{\unitlength}{\unitlength * \real{\svgscale}}%
    \fi%
  \else%
    \setlength{\unitlength}{\svgwidth}%
  \fi%
  \global\let\svgwidth\undefined%
  \global\let\svgscale\undefined%
  \makeatother%
  \begin{picture}(1,1)%
    \lineheight{1}%
    \setlength\tabcolsep{0pt}%
    \put(0,0){\includegraphics[width=\unitlength,page=1]{cf-connection.pdf}}%
    \put(0.09618829,0.59475538){\color[rgb]{0,0,0}\makebox(0,0)[lt]{\lineheight{1.25}\smash{\begin{tabular}[t]{l}$f$\end{tabular}}}}%
    \put(0.13799157,0.24016401){\color[rgb]{0,0,0}\makebox(0,0)[lt]{\lineheight{1.25}\smash{\begin{tabular}[t]{l}$g$\end{tabular}}}}%
    \put(0.52493315,0.12814771){\color[rgb]{0,0,0}\makebox(0,0)[lt]{\lineheight{1.25}\smash{\begin{tabular}[t]{l}$h$\end{tabular}}}}%
    \put(0.79442991,0.33305392){\color[rgb]{0,0,0}\makebox(0,0)[lt]{\lineheight{1.25}\smash{\begin{tabular}[t]{l}$i$\end{tabular}}}}%
    \put(0.76324844,0.68607262){\color[rgb]{0,0,0}\makebox(0,0)[lt]{\lineheight{1.25}\smash{\begin{tabular}[t]{l}$j$\end{tabular}}}}%
    \put(0.3868439,0.82861641){\color[rgb]{0,0,0}\makebox(0,0)[lt]{\lineheight{1.25}\smash{\begin{tabular}[t]{l}$k$\end{tabular}}}}%
    \put(0.50081844,0.53389163){\color[rgb]{0,0,0}\makebox(0,0)[lt]{\lineheight{1.25}\smash{\begin{tabular}[t]{l}$c$\end{tabular}}}}%
    \put(0.32346783,0.57121717){\color[rgb]{0,0,0}\makebox(0,0)[lt]{\lineheight{1.25}\smash{\begin{tabular}[t]{l}$\mathbf{cf}$\end{tabular}}}}%
    \put(0.31354265,0.41083204){\color[rgb]{0,0,0}\makebox(0,0)[lt]{\lineheight{1.25}\smash{\begin{tabular}[t]{l}$\mathbf{cg}$\end{tabular}}}}%
    \put(0.54932664,0.28301069){\color[rgb]{0,0,0}\makebox(0,0)[lt]{\lineheight{1.25}\smash{\begin{tabular}[t]{l}$\mathbf{ch}$\end{tabular}}}}%
    \put(0.61389758,0.46910332){\color[rgb]{0,0,0}\makebox(0,0)[lt]{\lineheight{1.25}\smash{\begin{tabular}[t]{l}$-\mathbf{cf}$\end{tabular}}}}%
    \put(0.5965737,0.54129418){\color[rgb]{0,0,0}\makebox(0,0)[lt]{\lineheight{1.25}\smash{\begin{tabular}[t]{l}$-\mathbf{cg}$\end{tabular}}}}%
    \put(0.48318081,0.66124097){\color[rgb]{0,0,0}\makebox(0,0)[lt]{\lineheight{1.25}\smash{\begin{tabular}[t]{l}-$\mathbf{ch}$\end{tabular}}}}%
  \end{picture}%
\endgroup%

%% file: sections/numerical-results.tex
The following sections show the improvements achieved in terms of curvature approximation, surface tension calculation and pressure velocity coupling on standard validation cases found in the literature. Because the quality of the Front is improved significantly by smoothing, the reconstruction of the interface can be avoided for small interface deformations.

Research data containing the numerical results are publicly available for: vector field reconstruction in OpenFOAM \footnote{\url{http://dx.doi.org/10.25534/tudatalib-61}}, SAAMPLE algorithm  data for the stationary droplet and low amplitude oscillation, \footnote{\url{http://dx.doi.org/10.25534/tudatalib-62}}, and the validation of the SAAMPLE algorithm with large amplitude droplet oscillations \footnote{\url{http://dx.doi.org/10.25534/tudatalib-136}}.

\subsection{Curvature approximation}
\label{subsection:curvature-results}
Due to the important role of curvature approximation for the computation of surface tension the accuracy of the techniques described in \cref{subsec:curvatureapp} is investigated here.
The following test setup has been used to obtain the results presented in this section:
A cubic domain $\Omega : [0,0,0]\times[4,4,4]$ is used, discretized with $n \in [16, 32, 64, 128]$ cells in each spatial direction. Two interface geometries, a sphere with radius $R=1$ and an ellipsoid with semi-axes $\Semiaxes = [3/2, 1, 1/2]$, are employed.
While the radius / semi-axes are kept constant, the interface centroids are generated randomly in a box-shaped region. The center of this region coincides with the center of $\Omega$ and its size is chosen such that the narrow band does not touch or intersect the domain boundary.
To examine the influence of the signed distance calculation, both the exact signed distance and the signed distance computed from the front are used in different setups. Accuracy of the different curvature models is evaluated with two norms of the relative curvature error
\begin{equation}
    L_\infty(e_{\MeanCurvature,\text{rel}}) =
     \underset{i}{\Max}\left( \frac{\Norm{\MeanCurvature_i - \ExactCurvature}}{\Norm{\ExactCurvature}} \right)
     \label{eq:linf-curvature}
\end{equation}
and 
\begin{equation}
    L_2(e_{\MeanCurvature,\text{rel}}) =
    \sqrt{\frac{1}{m}
    \sum_i\left(\frac{\Norm{\MeanCurvature_i - \ExactCurvature}}{\Norm{\ExactCurvature}}\right)^2}
    \label{eq:l2-curvature}
\end{equation}
where index $i$ denotes all cell faces at which the surface tension is evaluated
and $m$ is the number of such faces. We have chosen the face centers as evaluation location rather than the cell centers since the surface tension is discretized at the cell faces. Each setup (resolution, interface shape, signed distance calculation procedure, curvature model) is repeated 20 times with random placement as described above.

\begin{figure}
    \centering
    \begin{subfigure}[b]{0.48\textwidth}
        \includegraphics[width=\textwidth]{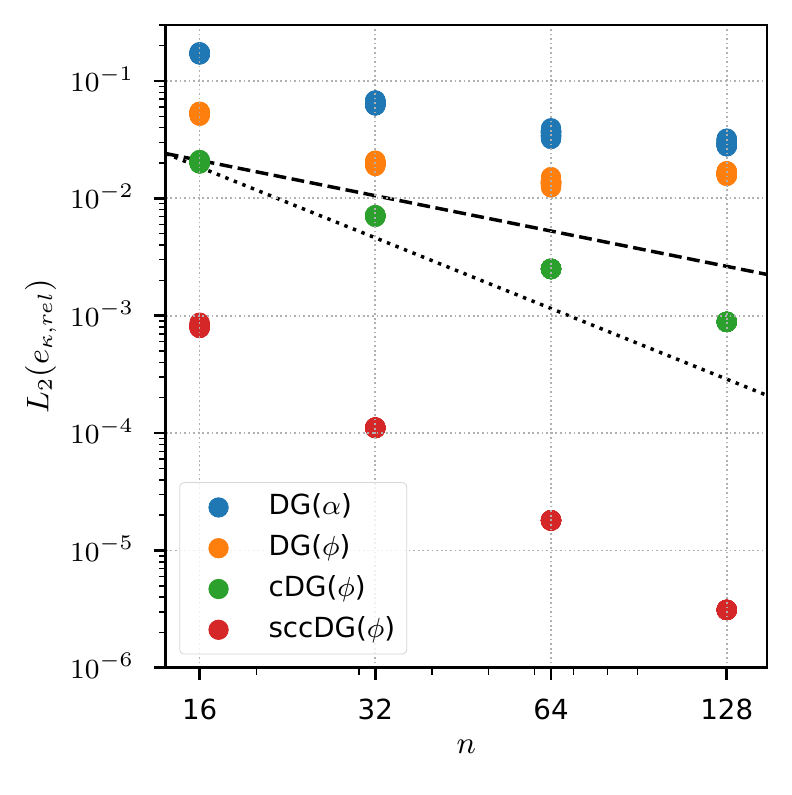}
        \caption{Sphere $R=1$, exact $\SignedDistance$}
        \label{plot:curvature-l2-00}
    \end{subfigure}
    ~
    \begin{subfigure}[b]{0.48\textwidth}
        \includegraphics[width=\textwidth]{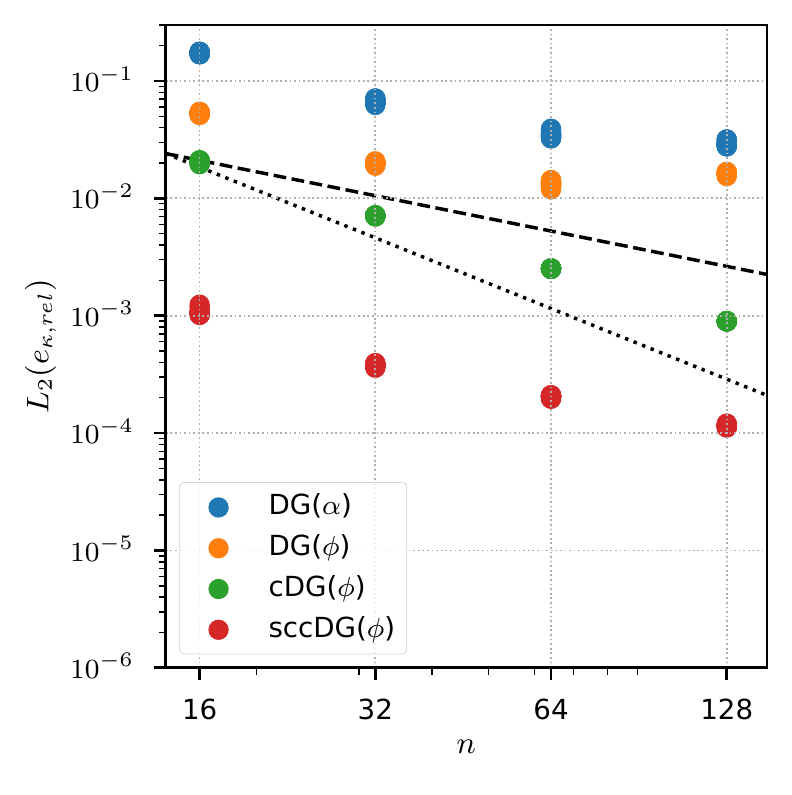}
        \caption{Sphere $R=1$, $\SignedDistance=f(\Front)$}
        \label{plot:curvature-l2-01}
    \end{subfigure}
    ~
    \begin{subfigure}[b]{0.48\textwidth}
        \includegraphics[width=\textwidth]{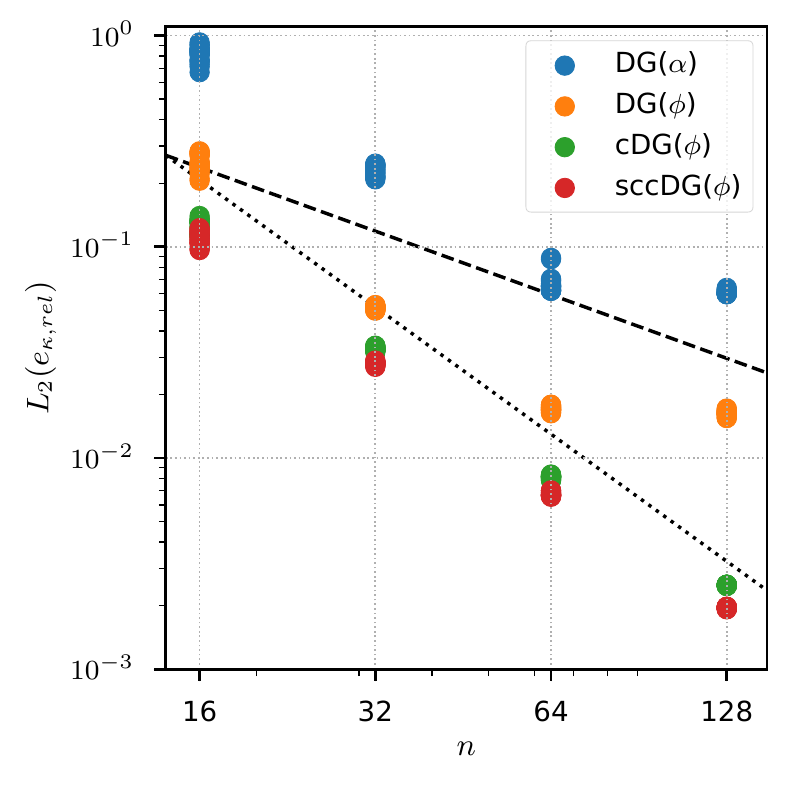}
        \caption{Ellipsoid $\mathbf{s}=(1.5, 1.0, 0.5)$, exact $\SignedDistance$}
        \label{plot:curvature-l2-20}
    \end{subfigure}
    ~
    \begin{subfigure}[b]{0.48\textwidth}
        \includegraphics[width=\textwidth]{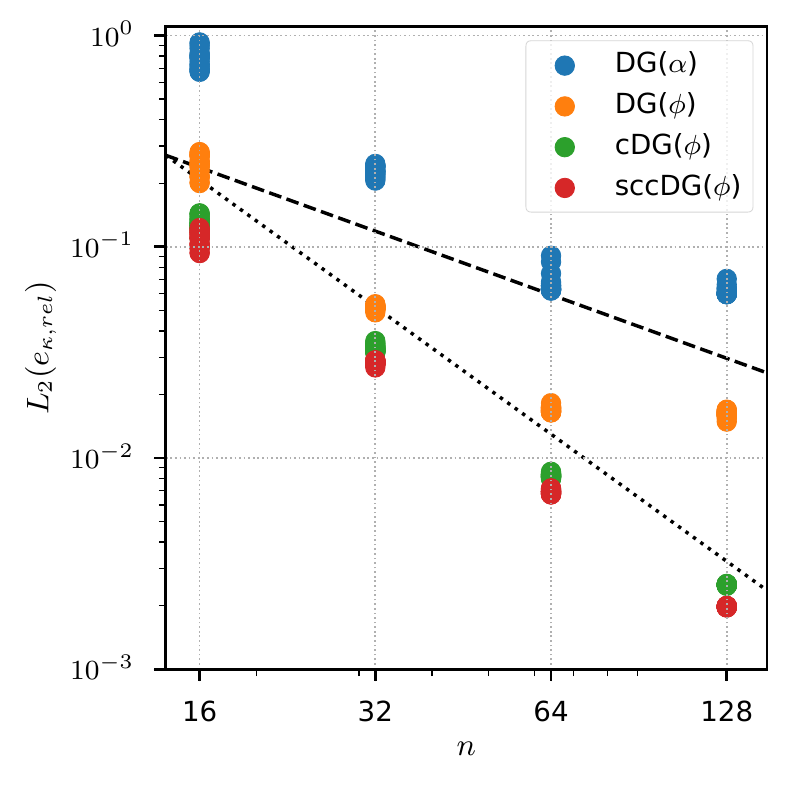}
        \caption{Ellipsoid $\mathbf{s}=(1.5, 1.0, 0.5)$, $\SignedDistance=f(\Front)$}
        \label{plot:curvature-l2-21}
    \end{subfigure}
    \caption{Scatter plots showing the $L_2$ norm of the relative curvature error for a sphere (upper row) and an ellipsoid (lower row) for different curvature models. The exact signed distance (left column) and the signed distance computed from the front (right column) have been used as input. A cubic domain $\Omega : [0,0,0] \times [4,4,4] $ with $n$ cells in each spatial direction. Each setup has been simulated 20 times by setting the centroid of the interface at a random position around the domain center. The dashed/dotted line indicates first/second order of convergence.
    }
    \label{fig:curvature-l2-comparison}
\end{figure}

The results for the $L_2$-norm are depicted as scatter plots in \cref{fig:curvature-l2-comparison} and the four different configurations are summarized in \cref{tab:curvature-configurations}.

\begin{table}[]
    \centering
    \begin{tabular}{r|ccc}
        acronym        & input field        & compact calculation & spherical correction \\
        \hline
        \dgMarkerfield & $\PhaseIndicator$  & no                  & no \\
        \dgSigneddistance & $\SignedDistance$ & no                & no \\
        \compactdgnc     & $\SignedDistance$ & yes                & no \\
        \compactsphere   & $\SignedDistance$ & yes                & yes
    \end{tabular}
    \caption{Acronyms of the tested curvature model configurations. Compact calculation refers to the modfication described in \cref{subsec:compactcurvature} and spherical correction to \cref{eq:spherical_correction}.}
    \label{tab:curvature-configurations}
\end{table}
Several conclusions can be drawn from these plots. First of all, scattering range is small compared to the error differences between different resolutions and different models. The only exception are the \compactdgnc{} and the \compactsphere{} model for the ellipsoidal interface where there is an overlap of data points.
Furthermore, the accuracy of the models increases in the order as they are listed in \cref{tab:curvature-configurations} for all setups.
While the signed distance as input field significantly improves the accuracy compared to the phase indicator field, the qualitative behavior is left unchanged: both models show convergence of decreasing order up to $n=64$. Further increase of the resolution does not reduce the $L_2$-norm (\dgMarkerfield{}) or reveals the onset of divergence (\dgSigneddistance{}).

Applying the compact curvature calculation idea of \citep{ShinJuric2009} described in \cref{subsec:compactcurvature} yields higher absolute accuracy and consistent convergence behavior. Order of convergence lies in-between one and two for a sphere which agrees with the estimation of the maximum error \cref{eq:relative_curvature_error_n}.

For the spherical correction approach \cref{eq:spherical_correction}, two observations can be made. First, there is no negative impact on the accuracy when applied to the non-spherical, ellipsoidal interface. Second, as can be expected for a spherical interface, the errors are reduced by an order of magnitude or more compared to the compact calculation without correction.

Finally, comparison of \cref{plot:curvature-l2-00} and \cref{plot:curvature-l2-01}
shows the impact of the signed distance calculation on the curvature calculation when the
\compactsphere{} model is used. 
For the curvature models \dgMarkerfield{} and \compactsphere{} which are used for the hydrodynamic test cases the results are summarized in \cref{tab:curvature-errors-exact-distance} and \cref{tab:curvature-errors-front-distance}. They illustrate the distinct improvement of both the $L_2$- and $L_\infty$-norm compared to the previous publication \cite{Maric2015}.

\begin{table}[]
    \centering
    \include{tables/curvatureErrors-exactDistance}
    \caption{Mean error norms ($\overline{L_\infty}$ and $\overline{L_2}$) of the curvature and their standard deviation $\sigma$ for the \dgMarkerfield{} and \compactsphere{} model using an exact signed distance. Between resolutions, the order of convergence is displayed.}
    \label{tab:curvature-errors-exact-distance}
\end{table}

\begin{table}[]
    \centering
    \include{tables/curvatureErrors-frontDistance}
    \caption{Mean error norms ($\overline{L_\infty}$ and $\overline{L_2}$) of the curvature and their standard deviation $\sigma$ for the \dgMarkerfield{} and \compactsphere{} model using a signed distance computed from the front. Between resolutions, the order of convergence is displayed.}
    \label{tab:curvature-errors-front-distance}
\end{table}

\subsection{Surface tension force reconstruction}
\label{subsec:result:forcerecon}

\begin{figure}[!hbt]
    \centering
    \begin{subfigure}[t]{.45\textwidth}
        \centering
        \includegraphics[]{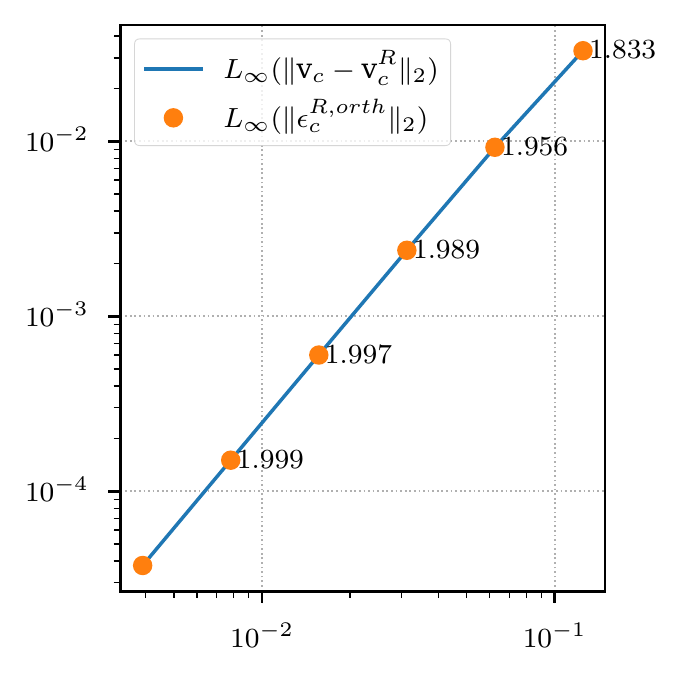}
        \caption{Convergence of the flux reconstruction for the "SHEAR 2D" velocity field.}
        \label{fig:recon:shear2D}
    \end{subfigure}
    \centering
    \begin{subfigure}[t]{.45\textwidth}
        \centering
        \includegraphics[]{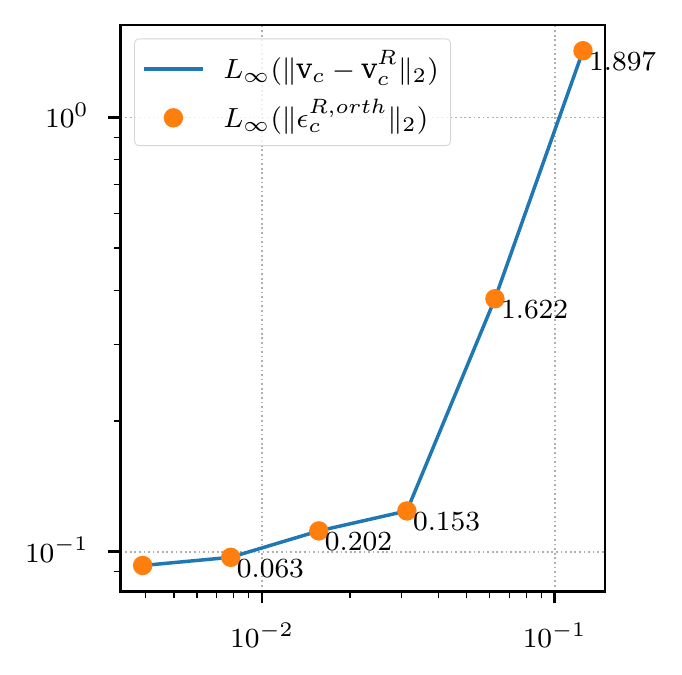}
        \caption{Convergence of the flux reconstruction for the "Hadamard-Rybczynsky" velocity field.}
        \label{fig:recon:hadamard}
    \end{subfigure}
    \caption{Convergence of the flux reconstruction.}
    \label{fig:recon}
\end{figure}

\Cref{fig:recon} contains results from two tests used to verify the second-order accuracy of the reconstruction operator described in \cref{subsec:num:forcerecon}. In both cases, velocity is reconstructed in cell centers, from the numerical scalar flux exactly defined at face centers. The first case, shown in \cref{fig:recon:shear2D}, is a single-phase solenoidal velocity field function known as the "single vortex test", often used to validate the advection of the fluid interface \citep{RiderKothe1998}. 

\begin{figure}[!hbt]
    \centering
    \includegraphics[width=0.6\textwidth]{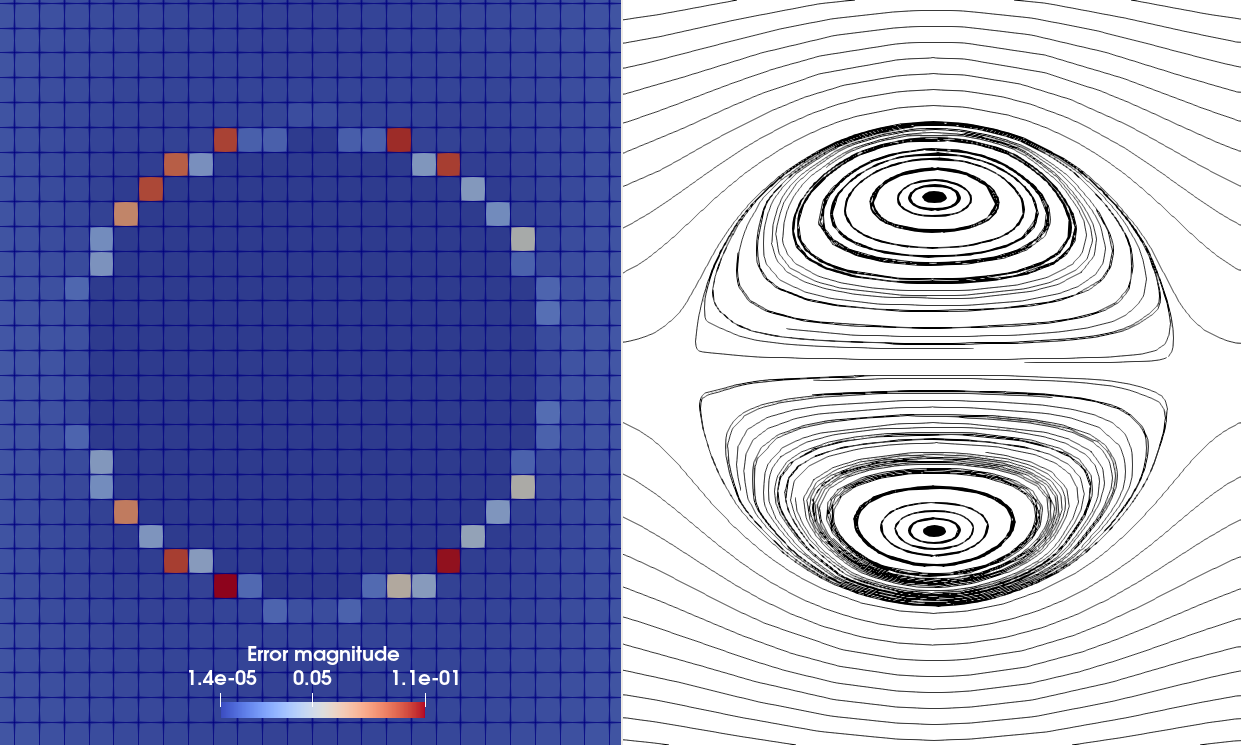}
    \caption{Error distribution for the Hadamard-Rybczynsky flow.}
    \label{fig:recon:hadamard:stream}
\end{figure}

For the other verification case, the creeping flow around a spherical interface is used, as given by the H\'{a}damard-Rybczynsky model. We have used velocity expressions available in \citep{Brenn2016}. The interface is a circle, of radius $R=0.15$, centered at $(0.5011, 0.507, 0.05)$. The outside viscosity is $\mu_o = 5\text{e-}04$ and the inside viscosity $\mu_i = 5\text{e-}03$ and the free stream velocity $\U_\infty=1.0$. \Cref{fig:recon:shear2D,fig:recon:hadamard} show that the reconstruciton error $\epsilon_c^{R,orth}$ given by \cref{eqn:recorth} exactly corresponds with the computed $L_\infty$ error norm $L_{\infty}(|\U_c - \U^R_c|)$. Therefore, \cref{eqn:recorth} can be used to verify the reconstruction operator convergence behavior both for single and two-phase flows. Second order accuracy is obtained only in the single-phase scenario, shown in \cref{fig:recon:shear2D}. However, as the discontinuity of the velocity gradient in the H\'{a}damard-Rybczynsky model increases with increasing mesh resolution, the convergence of the reconstruction operator deteriorates, as shown in cf. \cref{fig:recon:hadamard}. \Cref{fig:recon:hadamard:stream} confirms this, by showing the $L_\infty$ error norm distribution for the velocity field, where the error is concentrated at the fluid interface.

These results have an important consequence. Field reconstruction is performed by the SAAMPLE algorithm at the r.h.s. of the momentum equation for the surface tension force (together with other forces) as well as for the velocity field, at the end of the internal loop of the SAAMPLE algorithm. As clearly visible in \cref{fig:recon:hadamard:stream}, both the velocity field, and the surface tension force reconstruction introduce errors at the interface. Further improvements of the reconstruction operator are expected to improve the convergence and stability of the SAAMPLE algorithm and are left as future work. 

\subsection{Stationary droplet}
\label{subsection:stationary-droplet}
According to the Young-Laplace law, the velocity for a spherical droplet in equilibrium in the absence of gravity is $\Velocity = 0$ because the surface tension is balanced by the pressure jump across the interface. With a prescribed, constant curvature this case allows to test if a numerical method is \emph{well-balanced} \citep{Francois2006}. With a numerically approximated curvature, limitations of the numerical method with respect to the capillary number
\begin{equation}
    Ca = \frac{\Norm{\Velocity_\text{ref}}\Dynviscosity}{\Surfacetensioncoeff}
    \label{eq:capillary-number}
\end{equation}
due to so-called \emph{spurious currents} can be investigated. We adapt the setup used in \citep{Popinet2009,Abadie2015}. Our setup differs from those publications in two regards. First, it is three-dimensional instead of two-dimensional. Second, no symmetry is used: the complete droplet is simulated. The domain is $\Domain: [0,0,0]\times[1.6,1.6,1.6]$ with a spherical interface of $R=0.4$
centerd at $[0.800000012, 0.799999932, 0.800000054]$, to avoid exact overlap with mesh points.
The material properties are identical for the droplet and the ambient fluid with the density $\Density = 1$, the kinematic viscosity $\Kinviscosity = [8.165\text{e-2}, 2.582\text{e-2}, 8.165\text{e-3}, 0]$ and a surface tension coefficient of $\Surfacetensioncoeff = 1$. The values of $\Kinviscosity$ are chosen such that the Laplace number
\begin{equation}
    La = \frac{2R\Surfacetensioncoeff}{\Density \Kinviscosity^2}    
\end{equation}
assumes $La = [120, 1200, 12000, \infty]$. We prescribe Dirichlet boundary conditions for the pressure $\Pressure = 0 \text{ on }\Boundary$ and for the velocity $\Grad{\Velocity} \cdot \mathbf{n} = 0 \text{ on }\Boundary$. The initial conditions are $\Pressure(t_0) = 0$ and $\Velocity(t_0) = 0$. A time step of $\Delta t = 0.5 \Delta t_\text{cw}$ is chosen where
\begin{equation}
    \Delta t_\text{cw} = \sqrt{\frac{\Density h^3}{\pi \Surfacetensioncoeff}}
    \label{eq:capillary-time-step}
\end{equation}
is the time step limit due to capillary waves according to \citep{Denner2015}.

\begin{figure}
    \centering
    \includegraphics[scale=0.5]{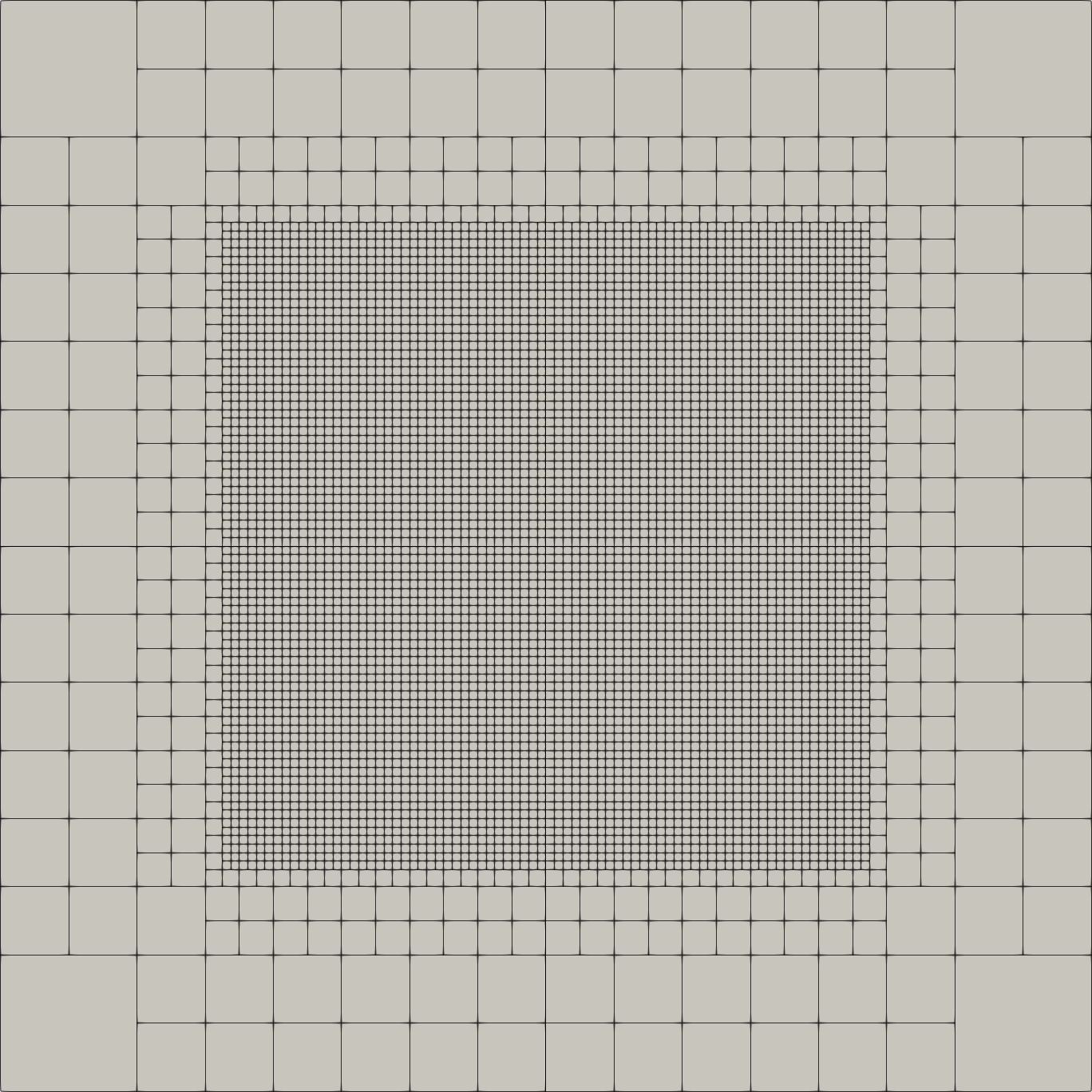}
    \caption{Cross-section of a 3D mesh used for the simulation of a stationary droplet with an equivalent resolution of $n_e=128$. Distribution of the cell sizes is chosen such that the interface is located within the finest, uniformly resolved region.}
    \label{fig:mesh-crosssection}
\end{figure}

For the spatial discretization hexahedral cells are used. To reduce the number of cells, the unstructured mesh is statically refined: small uniform cells are in the region of the narrow band, and larger cells are used away from the interface as shown in \cref{fig:mesh-crosssection}. To classify the mesh resolution we use a so-called equivalent resolution $n_e$. This is the number of cells along a spatial direction if the domain was resolved uniformly using the cell size in the interface region.

\subsubsection{Prescribed exact, constant curvature}
\label{subsec:constant-curvature-droplet}
To test if our numerical method is well-balanced, a stationary droplet is simulated with a prescribed, constant curvature.
\begin{figure}
    \centering
    \begin{subfigure}[b]{0.48\textwidth}
        \includegraphics[width=\textwidth]{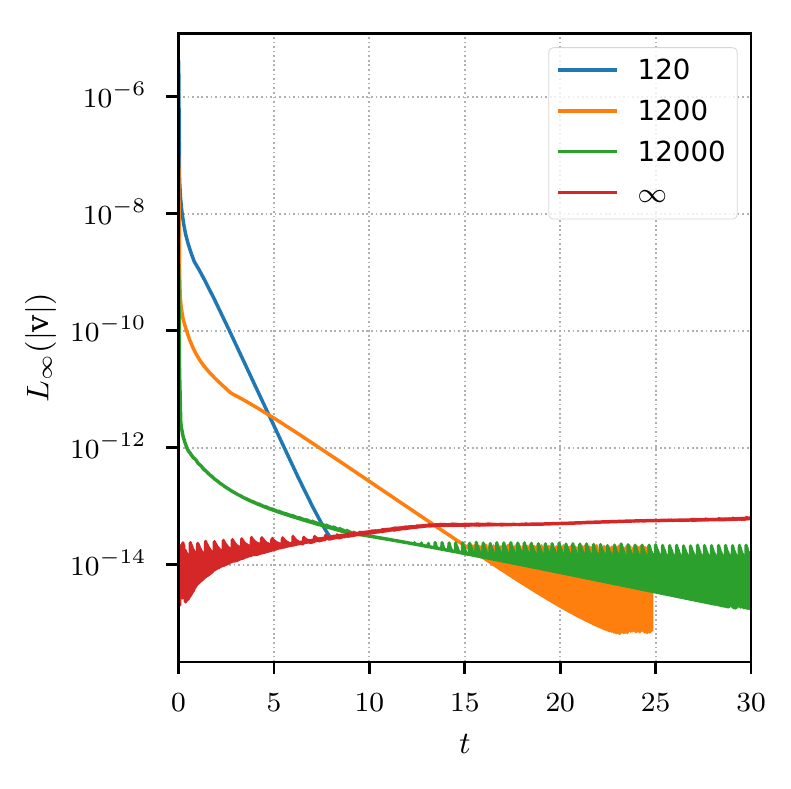}
        \caption{$n_e=16$}
        \label{fig:sd_exact_piso_n16}
    \end{subfigure}
    ~
    \begin{subfigure}[b]{0.48\textwidth}
        \includegraphics[width=\textwidth]{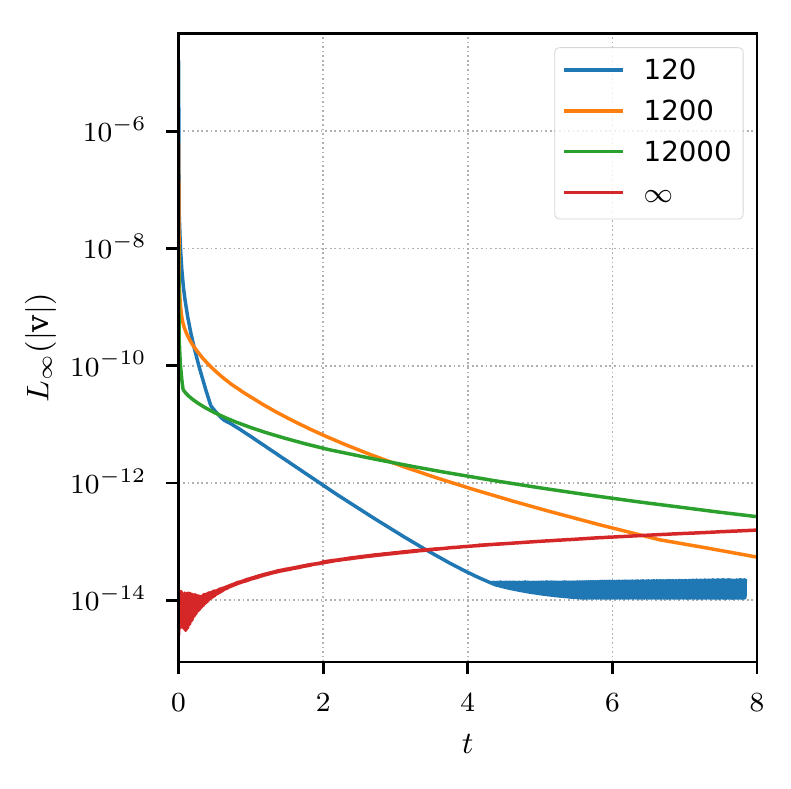}
        \caption{$n_e=64$}
        \label{fig:sd_exact_piso_n64}
    \end{subfigure}
    \caption{Temporal evolution of the spurious currents for the case of a stationary droplet for different Laplace numbers. The LENT method is configured as in \citep{Maric2015} using the PISO algorithm  and exact curvature. On the left, the results for an equivalent resolution of $n_e=16$ are displayed while the right graph has been computed with an equivalent resolution of $n_e=64$.}
    \label{fig:sd-exact-curvature-piso}
\end{figure}
Figure \ref{fig:sd-exact-curvature-piso} shows the temporal evolution of the spurious currents when the PISO algorithm is used to solve the pressure velocity system. Except for the inviscid case, the PISO algorithm does not obtain the equilibrium state within the first time step. Instead, there is a transient phase before the velocity magnitude falls below the linear solver tolerance.
This behavior can be understood by reformulating \cref{eq:pressure-correction} as
%
\begin{equation}
    \div{\left[\Dop^\Velocity_\Cell(\nabla \Pressure')_\Cell\right]} = 
        \div{\Velocity^*}
        + \underline{\div{\Hop_\Cell[\Velocity']}}
    \label{eq:pressure-correction-alternative}
\end{equation}
where $\Pressure' = \Pressure^* - \Pressure^\text{prev}$ is a correction to the old pressure field $\Pressure^\text{prev}$. Again, the underlined term $\underline{\div{\Hop_\Cell[\Velocity']}}$ is neglected. Thus, from \cref{eq:pressure-correction-alternative} it is clear that the pressure correction is driven by the divergence of the perliminary velocity field $\Velocity^*$. So, starting with a constant pressure field the only acting forces in the solution of
\cref{eq:momentum-predictor} are surface tension and viscous forces. The latter counteracts surface tension,
thus the resulting force is lower than in the inviscid case. One can then expect the volume defect also to be smaller than in the inviscid case. However, since the volume defect is the only source term for \cref{eq:pressure-correction-alternative}, the gradient of the updated
pressure field $\Pressure^* = \Pressure^\text{prev} + \Pressure'$ does not balance surface tension. Consequently, $\Linf{\Norm{\Velocity^**}} > \Tollinsolver$ can be expected after the explicit velocity update \cref{eq:velocity-update}.
Increasing the number of pressure correction itertions reduces $\Linf{\Norm{\Velocity(t=\Delta t)}}$, but it may require a considerable number of iterations to reach a given threshold as displayed in \cref{tab:sd-piso-iterations}.
\begin{table}[]
    \centering
    \include{tables/sd-piso-iterations}
    \caption{Number of pressure-correction iterations required to obtain $\Linf{\Norm{\Velocity}} < \text{1e-13}$ within the first time step for the stationary droplet with exact curvature. Results are shown for different Laplace numbers and two mesh resolutions.}
    \label{tab:sd-piso-iterations}
\end{table}
\begin{table}[]
    \centering
    \include{tables/stationary_droplet_exact_curvature_PISO_LSC}
    \caption{Spurious currents of the stationary droplet using the PISO and \Saample{} algorithm with exact curvature. Results are shown for different Laplace numbers $La$ and mesh resolutions $n_e$. Magnitude of the spurious currents is given after one time step and at the end of the simulations.}
    \label{tab:sd-exact-curvature}
\end{table}
This behavior motivated the development of the accuracy controlled \Saample{}  \cref{alg:pu-coupling}. Table \ref{tab:sd-exact-curvature} compares $\Linf{\Norm{\Velocity}}$ of PISO and \Saample{} after the first time step and at the end of simulation. For all configurations, the \Saample{} algorithm maintains $\Linf{\Norm{\Velocity}} < \Tollinsolver$ over the simulated time. This indicates that our method is balanced in the sense of \citep{Francois2006} and that \Saample{} is a suitable segregated solution algorithm for two-phase flows.

\subsubsection{Numerically approximated curvature}
\label{subsec:numerical-curvature-droplet}
\begin{figure}
    \centering
    \begin{subfigure}[b]{0.48\textwidth}
        \includegraphics[width=\textwidth]{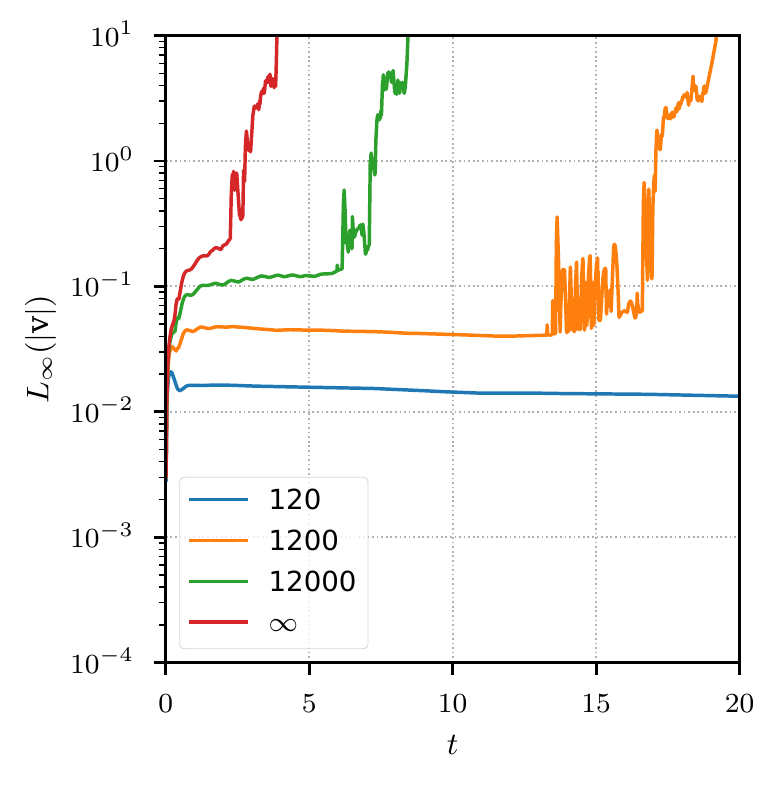}
        \caption{Resolution $n_e=16$, old state}
        \label{plot:sd-linf-n16-ref}
    \end{subfigure}
    ~
    \begin{subfigure}[b]{0.48\textwidth}
        \includegraphics[width=\textwidth]{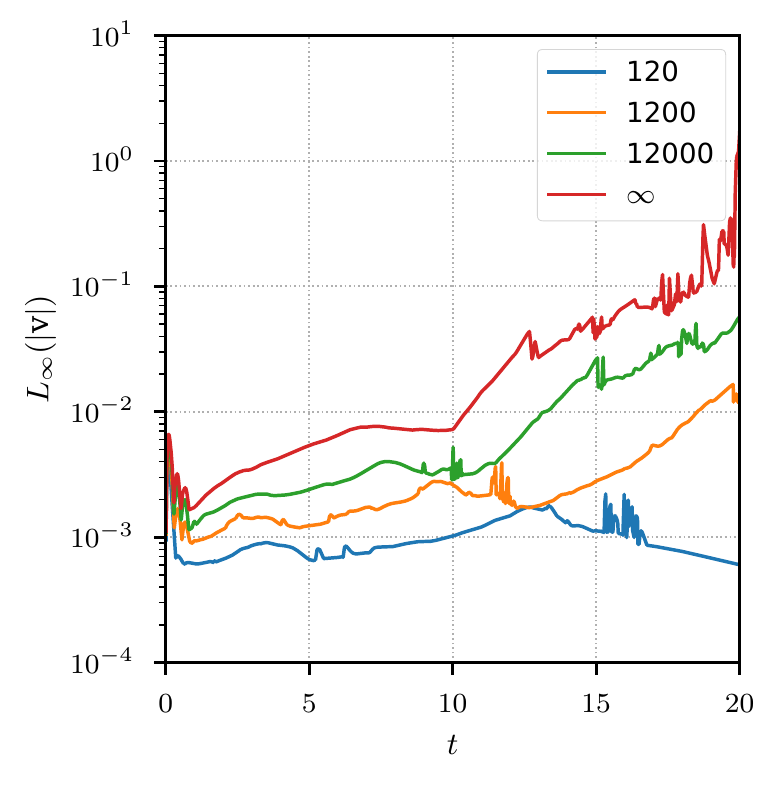}
        \caption{Resolution $n_e=16$, new state}
        \label{plot:sd-linf-n16-new}
    \end{subfigure}
    ~
    \begin{subfigure}[b]{0.48\textwidth}
        \includegraphics[width=\textwidth]{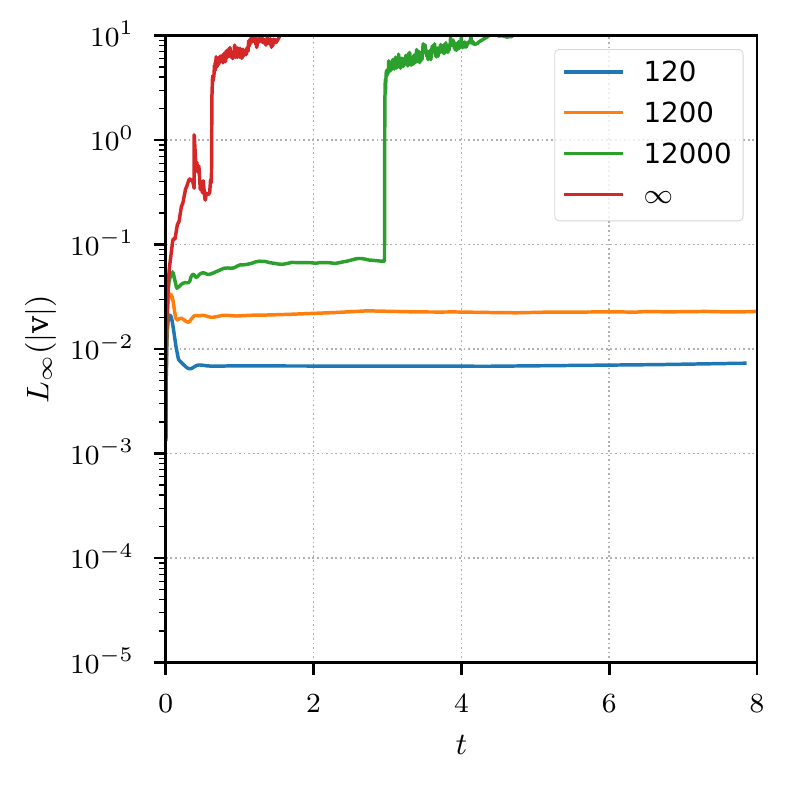}
        \caption{Resolution $n_e=64$, old state}
        \label{plot:sd-linf-n64-ref}
    \end{subfigure}
    ~
    \begin{subfigure}[b]{0.48\textwidth}
        \includegraphics[width=\textwidth]{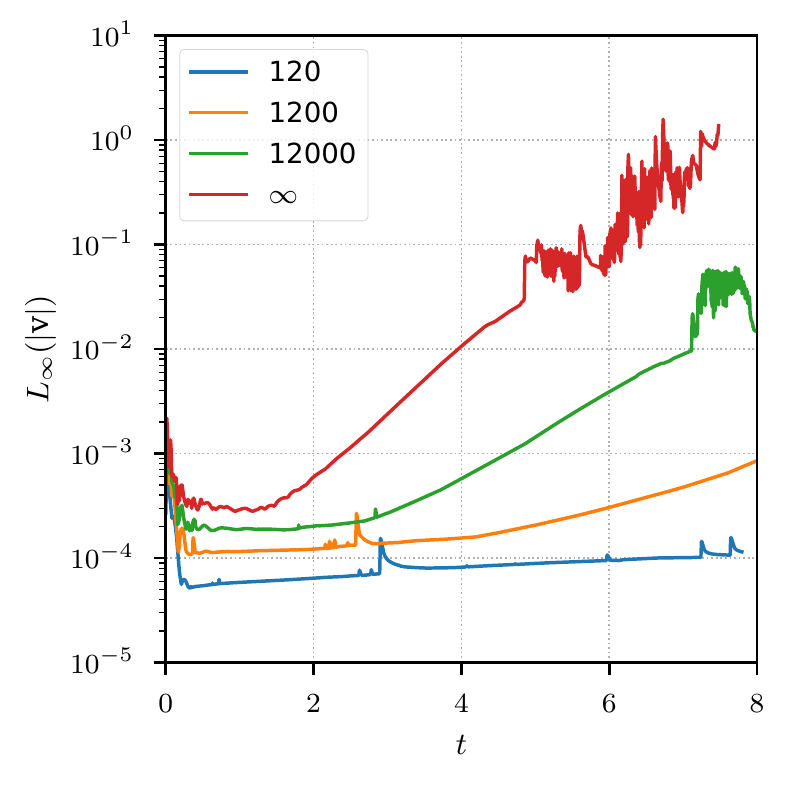}
        \caption{Resolution $n_e=64$, new state}
        \label{plot:sd-linf-n64-new}
    \end{subfigure}
    \caption{Evolution of spurious currents for the stationary droplet when using the LENT method. The left column shows the results obtained with the configuration from \citep{Maric2015}, the right column for the current configuration. In the upper row results for a resolution of $n_e = 16$ are displayed, in the lower row for $n_e = 64$. Each plot shows the results for different Laplace numbers.}
    \label{fig:stationary-droplet-linf-comparison}
\end{figure}

Two curvature models are used. For the LENT configuration from \citep{Maric2015} the
\dgMarkerfield{} model is used while the current configuration employs the \compactsphere{} model (see \cref{tab:curvature-configurations}). The results are compared in \cref{fig:stationary-droplet-linf-comparison} for two resolutions and four Laplace numbers. Overall, the new configuration of LENT reduces the spurious currents between one and two orders magnitude for the simulated time and Laplace numbers. With the old configuration \citep{Maric2015} simulations over the depicted time is only possible for $La=120$ ($n_e=16$) and $La=[120,1200]$ ($n_e=64$). Applying the modifications described in \cref{section:numerical-method} allows to simulate more physical time for all Laplace numbers.
%
\begin{figure}
    \centering
    \begin{subfigure}[b]{0.23\textwidth}
        \includegraphics[width=\textwidth]{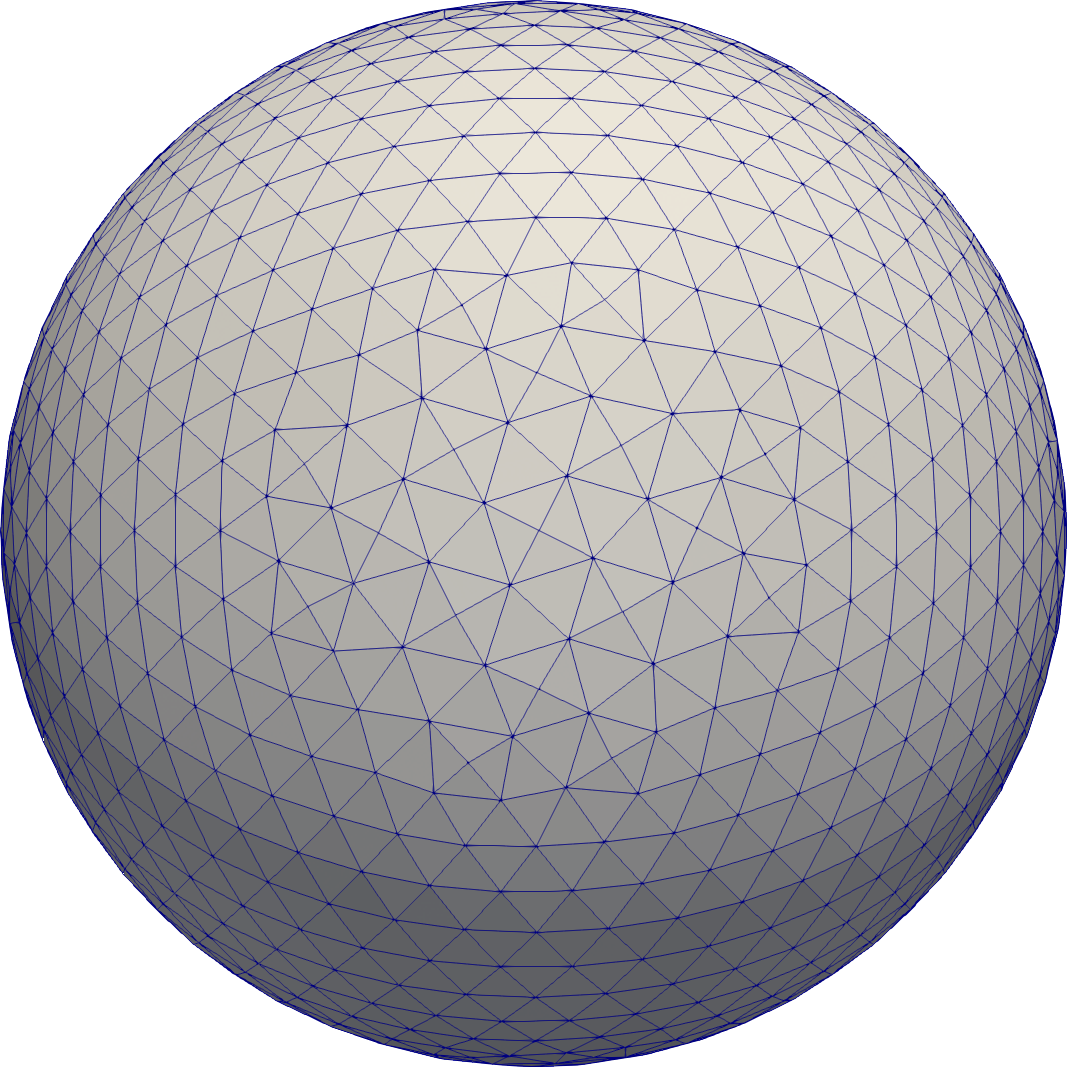}
        \caption{$\Front(t=0)$ (both)}
        \label{plot:sd-initial-front}
    \end{subfigure}
    ~
    \begin{subfigure}[b]{0.23\textwidth}
        \includegraphics[width=\textwidth]{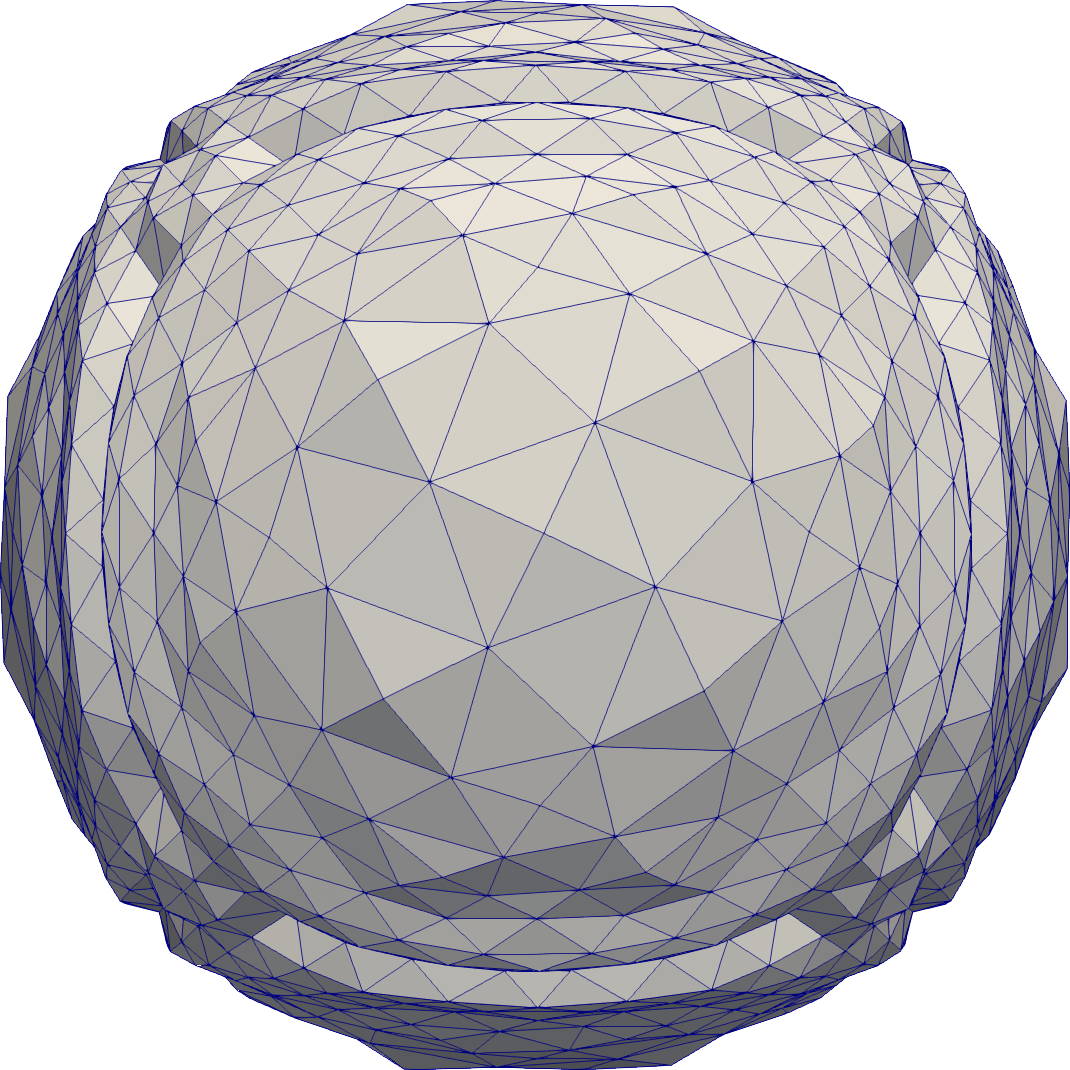}
        \caption{$\Front(t=20)$ (old)}
        \label{plot:sd-t20-old-lent}
    \end{subfigure}
    ~
    \begin{subfigure}[b]{0.23\textwidth}
        \includegraphics[width=\textwidth]{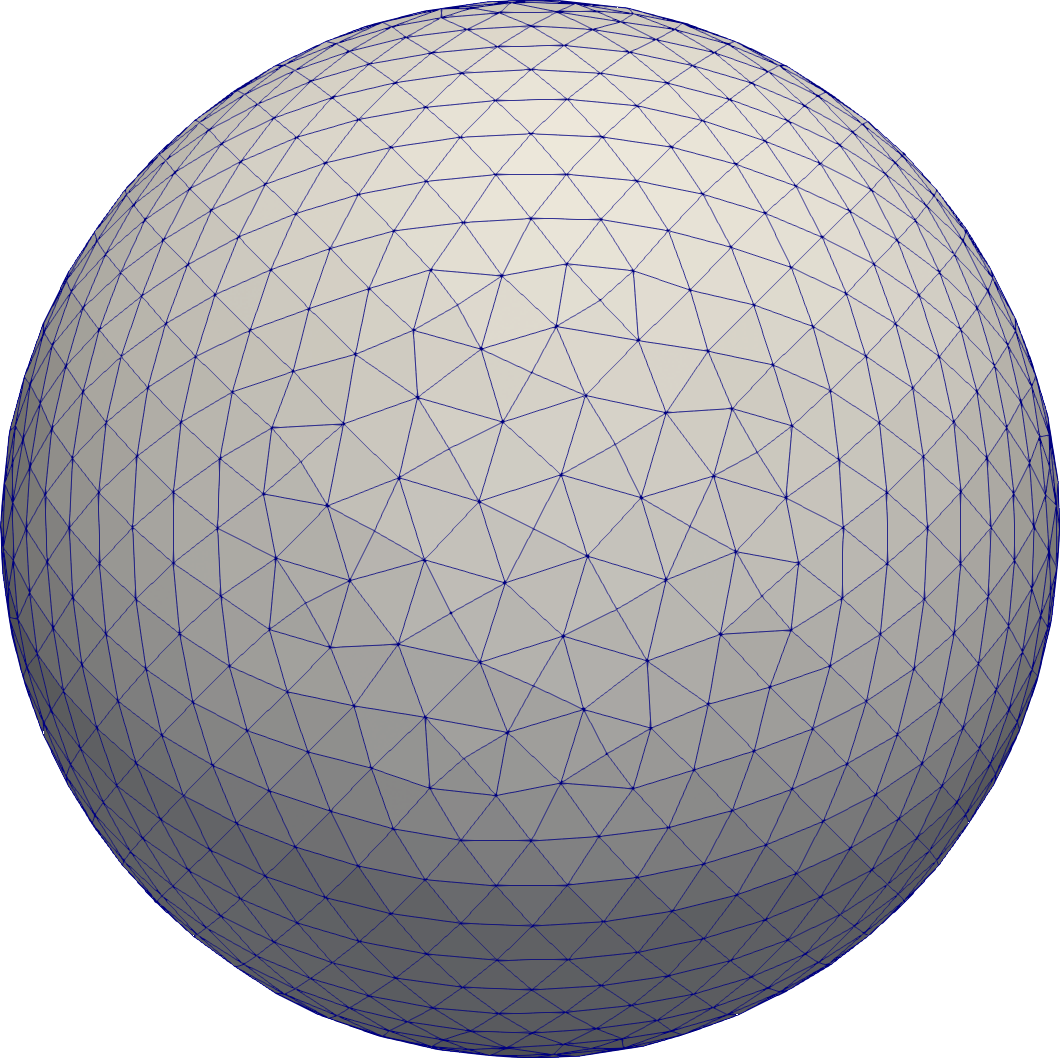}
        \caption{$\Front(t=20)$ (new)}
        \label{plot:sd-t20-new-lent}
    \end{subfigure}
    ~
    \begin{subfigure}[b]{0.23\textwidth}
        \includegraphics[width=\textwidth]{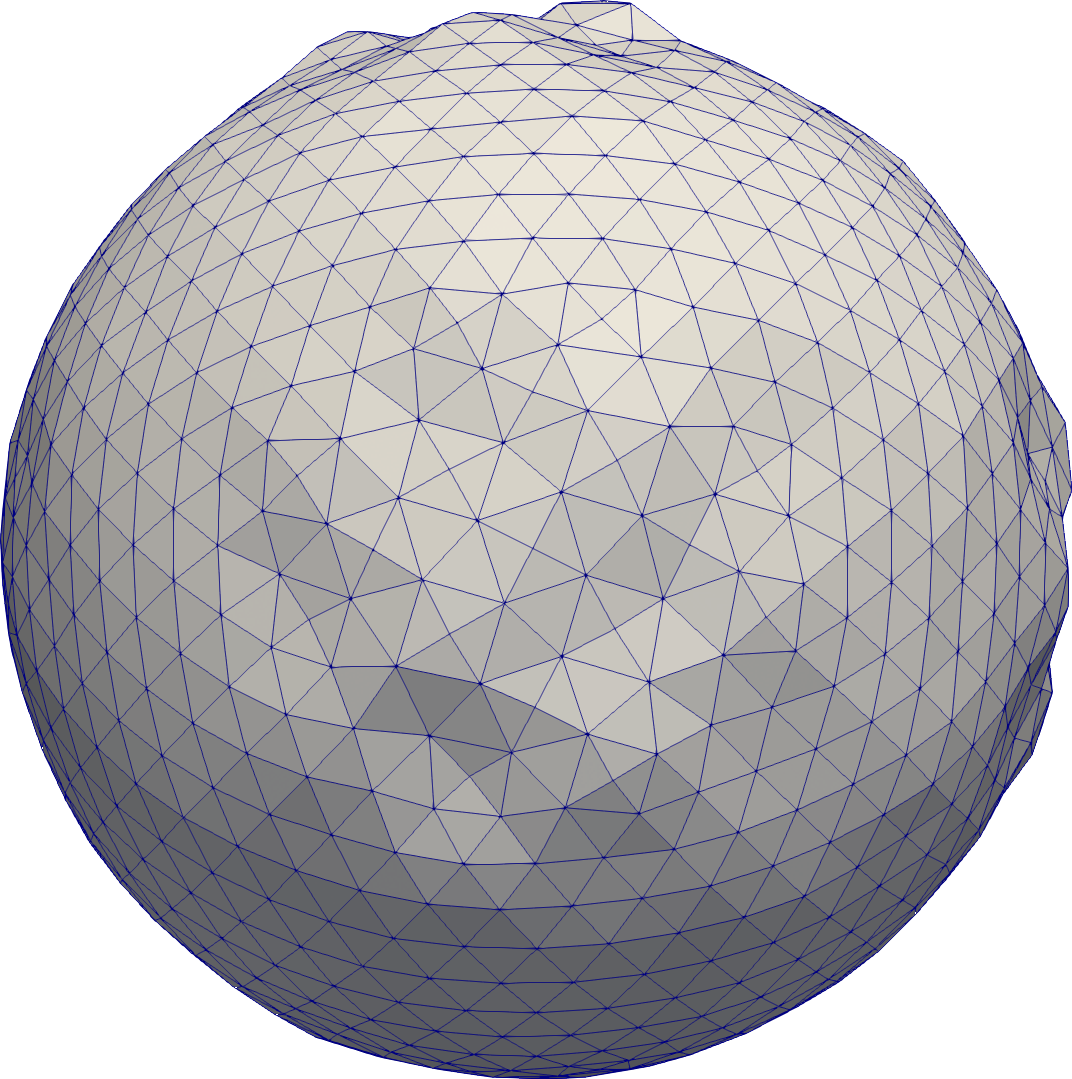}
        \caption{$\Front(t=40)$ (new)}
        \label{plot:sd-t40-new-lent}
    \end{subfigure}
    \caption{Front for the stationary droplet at different times for $La=120$. From left to right: initial front (identical for old and new configuration), front of the old configuration at $t=20$, front of the current configuration at $t=20$ and front of the current configuration at $t=40$.}
    \label{fig:sd-front-configurations}
\end{figure}
Yet, $\Linf{\Norm{\Velocity}}$ does not reach a quasi stationary state, but increases with time. The only expection from this is $La=120$ for the coarse resolution. A possible cause for this behavior might be the average number of front triangles per interface cell. For both resolutions, each interface cell contains 8-9 triangles on average, 
meaning that the front's resolution is notably finer than the resolution of the volume mesh. So, the relatively coarse resolution of the velocity field, which drives the motion of the front, may prevent that an equilibrium or quasi stationary state is reached. Instead, small scale perturbations accumulate in the vertex positions. These perturbations feed back through different parts of the algorithm (signed distance calculation, curvature approximation, surface tension) into the velocity. Over time, the perturbations become visible as shown in \cref{plot:sd-t40-new-lent}. Currently, it is not possible to change the average number of triangles per cell as this number is inherently linked to the interface reconstruction algorithm, whose improvement is left as future work. 

%

\subsection{Translating droplet}
\label{subsection:translating-droplet}
As pointed out by Popinet \citep{Popinet2009}, the solution to a stationary droplet also holds in a moving reference frame. Yet this variant is better suited to study the influence of interface advection as the droplet moves through the fixed cells of the mesh. Again, we adapt the two-dimensional setup from \citep{Popinet2009} for three spatial dimensions. Material properties are same as for the stationary droplet (\cref{subsection:stationary-droplet}). The radius of the droplet is $R=0.2$ and its center initially placed at $\mathbf{c}=[0.5, 0,5, 0.4]$ in $\Domain~:~[0, 0, 0] \times [5R, 5R, 6R]$. The constant, uniform background velocity is $\Velocity_\text{bg} = [0, 0, 1]$. As boundary conditions $\Grad{\Velocity} \cdot \mathbf{n} = 0$ and $\Pressure = 0$ is prescribed for the boundary part where $z=6R$, for the rest of the boundary $\Velocity = \Velocity_\text{bg}$ and $\partial \Pressure / \partial \mathbf{n} = 0$ is set. The initial conditions are $\Pressure(t_0) = 0$ and $\Velocity(t_0) = \Velocity_\text{bg}$. Simulation duration is chosen as $t=0.4$, so the that the droplet is advected by one diameter.

\begin{figure}
    \centering
    \begin{subfigure}[b]{0.48\textwidth}
        \includegraphics[width=\textwidth]{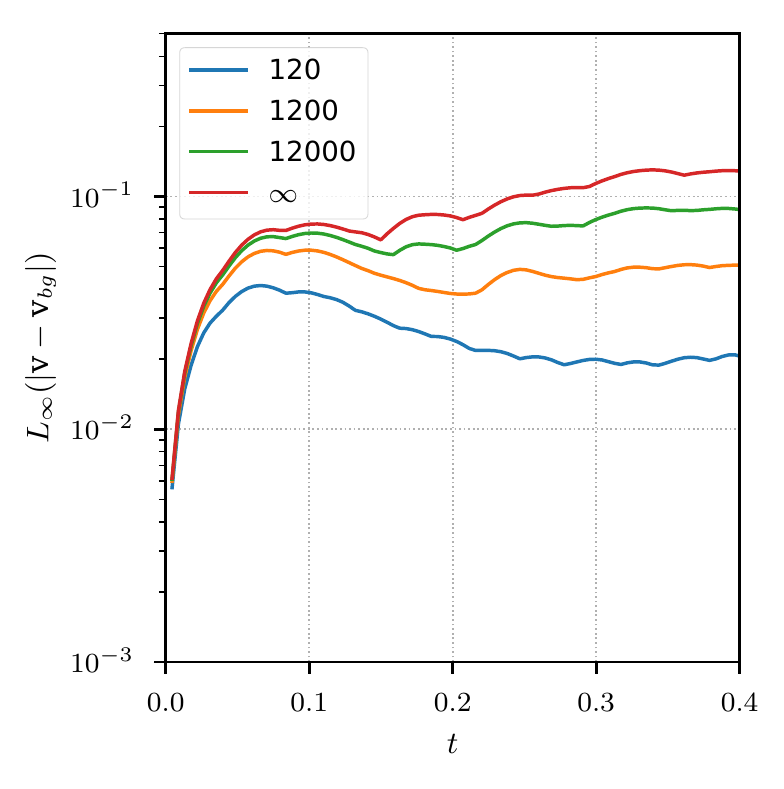}
        \caption{Resolution $n_e=16$, old state}
        \label{plot:td-linf-n16-ref}
    \end{subfigure}
    ~
    \begin{subfigure}[b]{0.48\textwidth}
        \includegraphics[width=\textwidth]{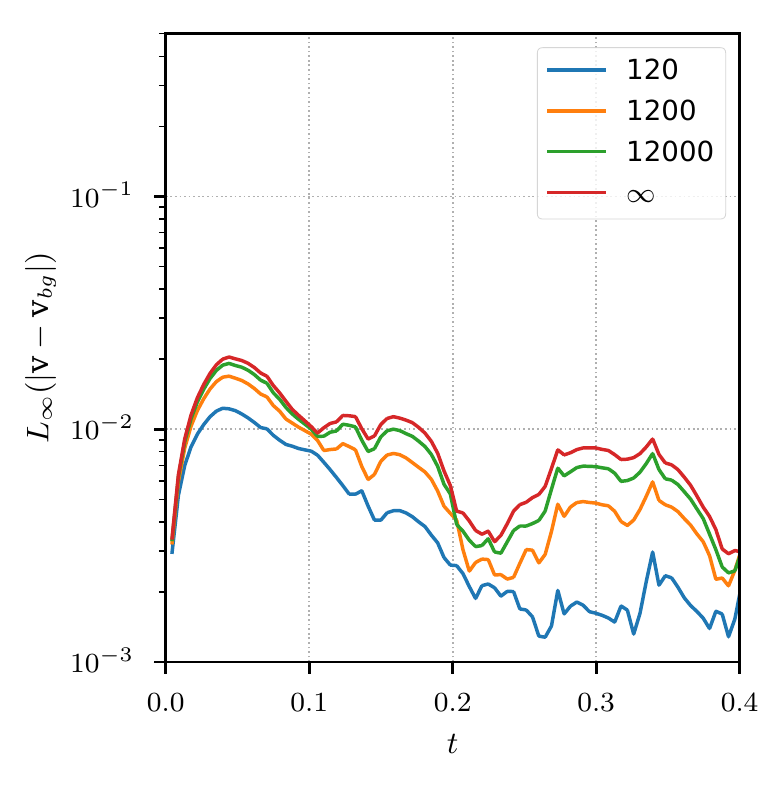}
        \caption{Resolution $n_e=16$, new state}
        \label{plot:td-linf-n16-new}
    \end{subfigure}
    ~
    \begin{subfigure}[b]{0.48\textwidth}
        \includegraphics[width=\textwidth]{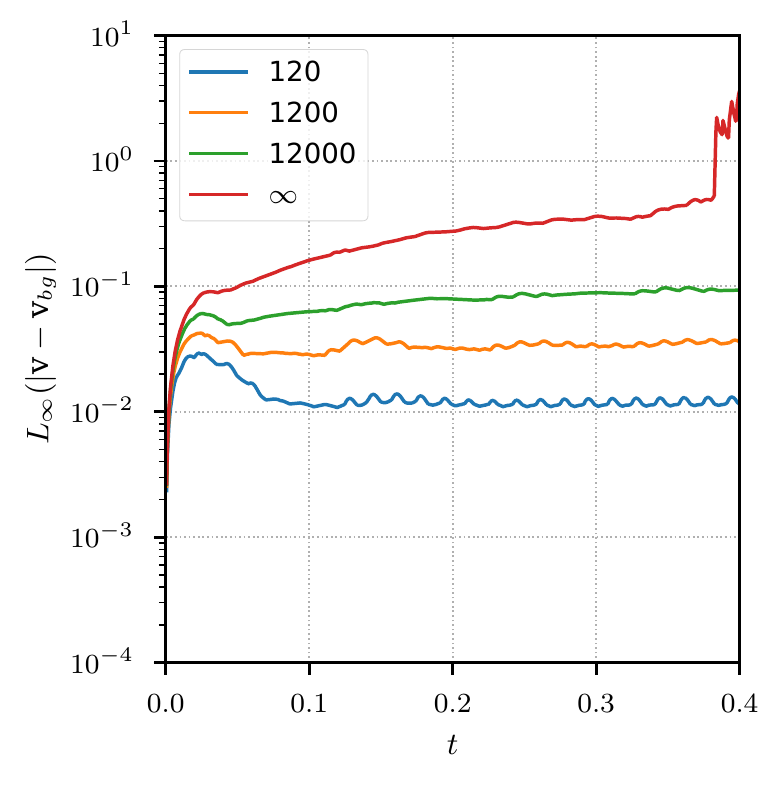}
        \caption{Resolution $n_e=64$, old state}
        \label{plot:td-linf-n64-ref}
    \end{subfigure}
    ~
    \begin{subfigure}[b]{0.48\textwidth}
        \includegraphics[width=\textwidth]{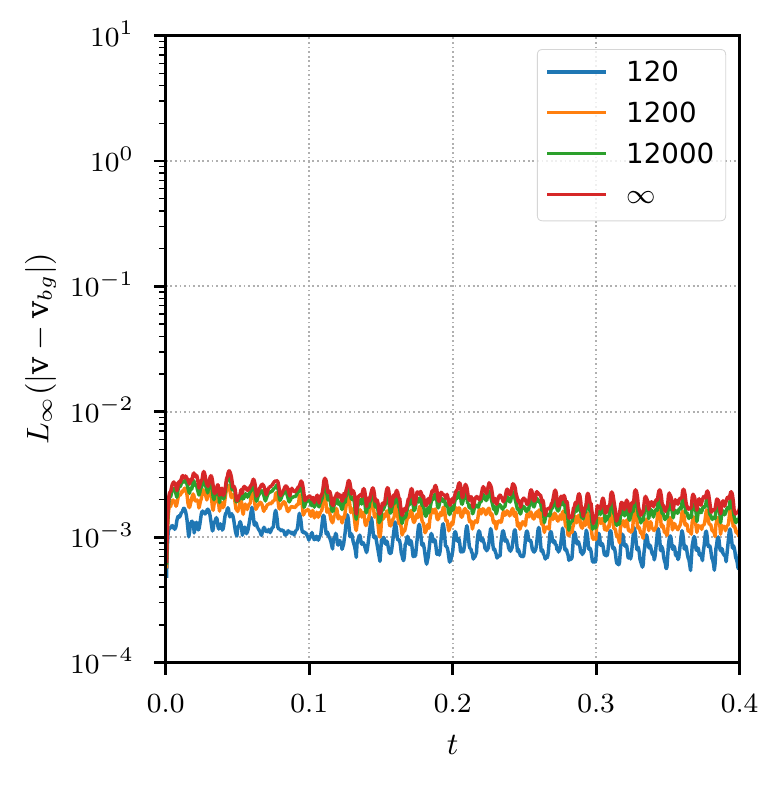}
        \caption{Resolution $n_e=64$, new state}
        \label{plot:td-linf-n64-new}
    \end{subfigure}
    \caption{Evolution of spurious currents for the translating droplet when using the LENT method. The left column shows the results obtained with the configuration from \citep{Maric2015}, the right column for the current configuration. In the upper row results for a resolution of $n_e = 16$ are displayed, in the lower row for $n_e = 64$. Each plot shows the results for different Laplace numbers.}
    \label{fig:translating-droplet-linf-comparison}
\end{figure}

In \cref{fig:translating-droplet-linf-comparison} the temporal evolution of the velocity deviations from the background velocity field $\Velocity_\text{bg}$ is displayed. As for the stationary droplet (\cref{subsection:stationary-droplet}), the figure compares two configurations of LENT for two resolutions. The improvements are similar to the stationary droplet with spurious currents reduced between one and two orders of magnitude. For $n_e=64$ and $La=[12000,\infty]$, the qualitative behavior changed also. The magnitude of spurious currents oscillates around its initial level while it increases for the previous configuration. Popinet \citep{Popinet2009} reports the period of the oscillations to scale with $\Norm{\Velocity}/h$ as the droplet moves through the cell layers of the mesh.\\
In a comprehensive comparison study Abadie et al. \citep{Abadie2015} show $Ca_\text{max}$ for different VoF and level set methods on structured meshes. They use the same parameters as in this publication ($n_e=64$, $La=12000$), albeit in a two-dimensional setting. The results are in the range $O(10^{-4}) < Ca_\text{max} < O(10^{-3})$ for the VoF methods and in the range $O(10^{-6}) < Ca_\text{max} < O(10^{-5})$ for the level set methods. With $\Norm{\Velocity}_\text{max} \approx 3\text{e-}3$ LENT maintains $Ca_\text{max} \approx 2.4\text{e-}5$, achieving more accurate results than the tested VoF methodes and comparable accuracy with regard to level set methods. In this case, the Lagrangian advection is advantageous as the movement of the front vertices due to $\Velocity_\text{bg}$ is captured exactly by first order spatial interpolation and first order temporal integration (\cref{eq:kinematic}). The errors arise from the signed distance calculation, influencing the calculation of $\PhaseIndicator$ and the approximation of $\MeanCurvature$.

\subsection{Oscillating droplet}
\label{subsection:oscillating-droplet}

\subsubsection{Comparison to analytic solution}
\label{subsection:od-small-amplitude}
\begin{figure}
    \centering
    \begin{subfigure}[b]{0.48\textwidth}
        \includegraphics[width=\textwidth]{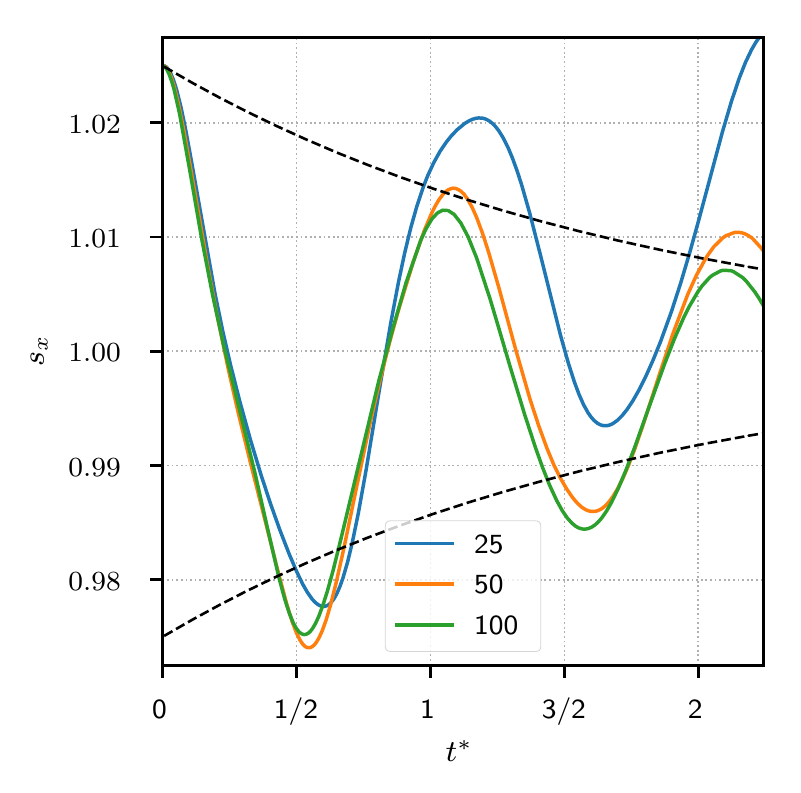}
        \caption{$\Kinviscosity = 0.05$, old state}
        \label{plot:od-sx-nuHigh-old}
    \end{subfigure}
    ~
    \begin{subfigure}[b]{0.48\textwidth}
        \includegraphics[width=\textwidth]{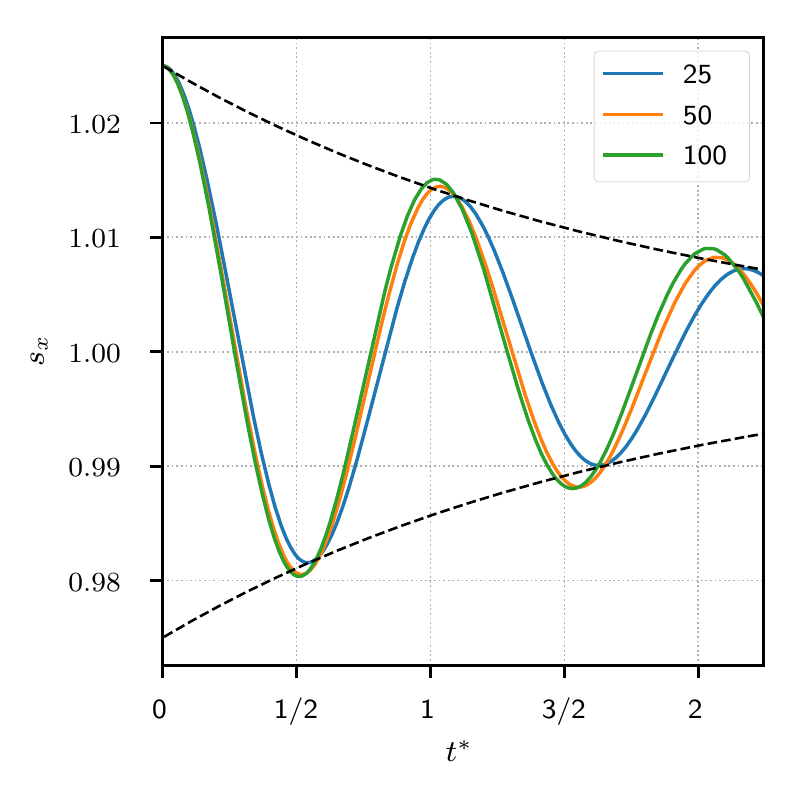}
        \caption{$\Kinviscosity = 0.05$, new state}
        \label{plot:sd-sx-nuHigh-new}
    \end{subfigure}
    ~
    \begin{subfigure}[b]{0.48\textwidth}
        \includegraphics[width=\textwidth]{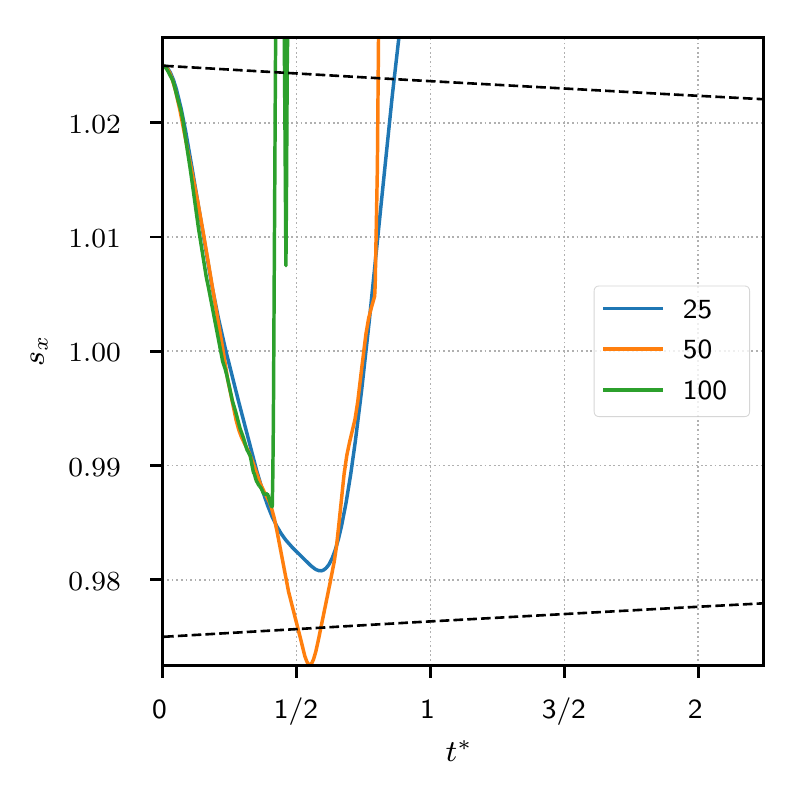}
        \caption{$\Kinviscosity = 0.005$, old state}
        \label{plot:od-sx-nuLow-old}
    \end{subfigure}
    ~
    \begin{subfigure}[b]{0.48\textwidth}
        \includegraphics[width=\textwidth]{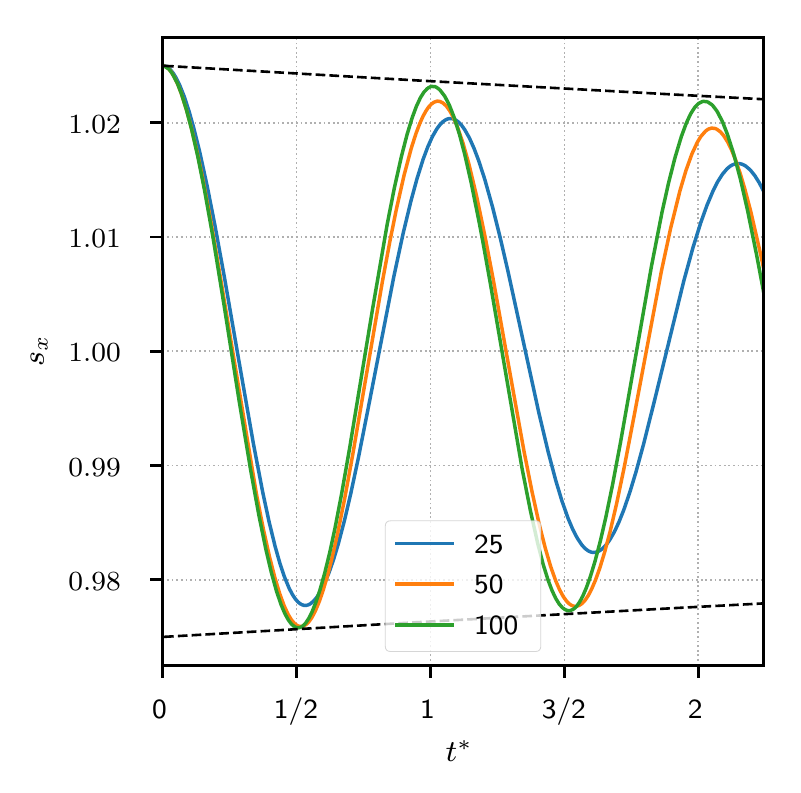}
        \caption{$\Kinviscosity = 0.005$, new state}
        \label{plot:sd-sx-nuLow-new}
    \end{subfigure}
    \caption{Temporal evolution of the $x$ semi-axis $\Semix$ for the oscillating droplet. Time is non-dimensionalized with the analytical period $T_2=2\pi/\ODfreq_2$. The dashed lines represent the exact envelope of $\Semix$ ($R_0 + a_2(t)$ and $R_0 - a_2(t)$, see \cref{eq:decay-rate}). Each plot shows the results for three mesh resolutions $n_e$. The case has been simulated using the LENT configuration from \cite{Maric2015} (left column) und the current configuration (right column) with different kinematic viscosities  of the droplet (upper and lower row).}
    \label{fig:oscillating-droplet-semiaxis-evolution}
\end{figure}

To analyze the accuracy of LENT with interface deformation
we adopt the setup of an oscillating droplet given in \citep{Shin2002,Shin2011}. For this case, Lamb derived an analytical solution. The oscillation frequency of an inviscid droplet is given by
\begin{equation}
    \ODfreq^2_n = \frac{n(n+1)(n-1)(n+2)\Surfacetensioncoeff}{
        [(n+1)\Density_d + n\Density_a]R_0^3}
    \label{eq:od-frequency}
\end{equation}
with the mode number $n$, the droplet density $\Density_d$, the density of the ambient fluid $\Density_a$ and the radius of the unperturbed droplet $R_0$. In case of a viscous fluid, the amplitude $a_n(t)$ decreases over time
\begin{equation}
    a_n(t) = a_0 e^{-\gamma t},\quad \gamma = \frac{(n-1)(2n+1)\Kinviscosity}{R_0^2}.
    \label{eq:decay-rate}
\end{equation}
The initial interface shape is
\begin{equation}
    R(\theta, t) = R_0 + \epsilon P_n(\cos{\theta})\sin{(\ODfreq_n t)},\quad \theta \in [0,2\pi],
    \label{eq:od-initial-interface}
\end{equation}
where $P_n$ denotes the $n$-th order Legendre polynom.\\
The domain is $\Domain : [0, 0, 0] \times [4, 4, 4]$, the interface is initialized with $R_0=1$, $n=2$, $\epsilon=0.025$ and $t = \pi/(2\ODfreq_n)$ with its center at $[2.00001, 1.99999, 2.0000341]$. Material parameters are $\Density_d = 10$, $\Density_a = 0.1$, $\nu_d = [0.05, 0.005]$, $\nu_a = 5\text{e-}4$ and $\Surfacetensioncoeff = 10$. Initial fields at $t=0$ are $\Velocity_0 = 0$ and $\Pressure_0 = 0$. Dirichlet boundary conditions are used for the pressure ($\Pressure=0$) and $\Grad{\Velocity} \cdot \mathbf{n} = 0$ for $\Velocity$. The semi-axis length is computed as
\begin{equation}
    \Semix = \frac{\underset{k}{\max}(\x^k_\Gamma \cdot \mathbf{e}_x) - \underset{k}{\min}(\x^k_\Gamma \cdot \mathbf{e}_x)}{2}
    \label{eq:semi-axis-computation}
\end{equation}
in each time step.\\
In \cref{fig:oscillating-droplet-semiaxis-evolution} the evolution of the semi-axis $\Semix$ is depicted for the previous configuration of LENT \citep{Maric2015} and the current one with two different kinematic viscosities. For $\Kinviscosity_d=0.05$, both configurations capture the qualitative behavior. However, the previous configuration shows considerable deviations with regard to the temporal evolution of $\Semix$. This can be attributed to development of spurious currents. First, the droplet deforms towards a cubic shape similar to \cref{plot:sd-t20-old-lent}, resulting in a smaller $\Semix$ around $t^*=1/2$ than analytically predicted. A second effect is that small wave like perturbations with wavelength comparable to the cell size $h$ grow over time. Since the displacement of a single vertex can already change the result of \cref{eq:semi-axis-computation}, $\Semix$ is considerably larger than the analytical prediction at later times, depending on the resolution. The setup reported in this publication, however, shows much smaller deviations from the analytical solution. While $\Semix$ decays a bit slower than predicted by \cref{eq:decay-rate}, the numerical period converges with mesh refinement.\\
Setting $\Kinviscosity=0.005$, the previous configuration is not able to simulate one oscillation. Due to decreased dissipation, perturbations of the front amplify themselves faster than for the more viscous setup and eventually lead to the crash of the simulation. With the new configuration, however, this setup becomes viable. Both amplitude and period converge with mesh resolution and at $n_e=100$ the numerical results agree very well with \cref{eq:od-frequency} and \cref{eq:decay-rate}.

%

\subsubsection{Comparison to experiment}
\label{subsection:od-large-amplitude}
The numerical results presented so far rely on analytical solutions for verification. In this section, we validate the proposed method against
experiments conducted by Trinh and Wang \citep{Trinh1982}. They investigate oscillations of droplets for which, in contrast to \cref{subsection:od-small-amplitude}, the amplitude cannot be considered
small compared to the equivalent radius of the droplet. For the experiments, single silicone oil drops
are suspended in water. Each drop is kept at a stable position using acoustic radiation pressure generated by an ultrasonic transducer. A second transducer
is used to drive the droplet oscillations. Besides forced oscillations, the authors investigate the damping of free large amplitude oscillations (section 5 in
\citep{Trinh1982}). In the following, we examine to what degree our LENT method is able to reproduce Trinh and Wang's experimental results.\\
The numerical setup is as follows. The fluid properties are as given in \citep{Trinh1982} with $\Density_a = 998~\text{kg}/\text{m}^3$, $\Kinviscosity_a = 0.95\text{e-}6~\text{m}^2/\text{s}$
for the ambient phase (water), $\Density_d = 1001~\text{kg}/\text{m}^3$, $\Kinviscosity_d = 3.2\text{e-}6~\text{m}^2/\text{s}$ for the droplet phase (silicone oil) and $\Surfacetensioncoeff = 0.037~\text{N}/\text{m}$.
A domain $\Domain : [0~\text{cm},0~\text{cm},0~\text{cm}] \times [8~\text{cm},8~\text{cm},8~\text{cm}]$ is used with equivalent resolutions of $n_e \in [64, 128, 256]$. The initial fields are
$\Pressure_0 = 0$ and $\Velocity_0 = 0$. Homogeneous Dirichlet boundary conditions are used for the pressure and
homogeneous Neumann boundary conditions for the velocity field. The interface is initialized as a prolate spheroid, centerd at $[4~\text{cm},4~\text{cm},4~\text{cm}]$ with two
semi-axes configurations:
$\Semiaxes_A = [8.02~\text{mm}, 5.46~\text{mm}, 5.46~\text{mm}]$ and $\Semiaxes_B = [9.18~\text{mm}, 5.1~\text{mm}, 5.1~\text{mm}]$. Configuration $A$ corresponds to
a droplet volume of $V_d = 1~\text{cm}^3$ and a semi-axes ratio of $\frac{L}{W}=1.47$, while configuration $B$ corresponds to
$V_d=1~\text{cm}^3$ and $\frac{L}{W}=1.80$ in \citep{Trinh1982}. The time step is set to $\Delta t = 0.5\Delta t_{cw}$ (\cref{eq:capillary-time-step}), giving
$\Delta t_{64} = 2~\text{ms}$, $\Delta t_{128} = 0.72~\text{ms}$ and $\Delta t_{256} = 0.26~\text{ms}$.
.
\begin{figure}
    \centering
    \begin{subfigure}[b]{0.48\textwidth}
        \includegraphics[width=\textwidth]{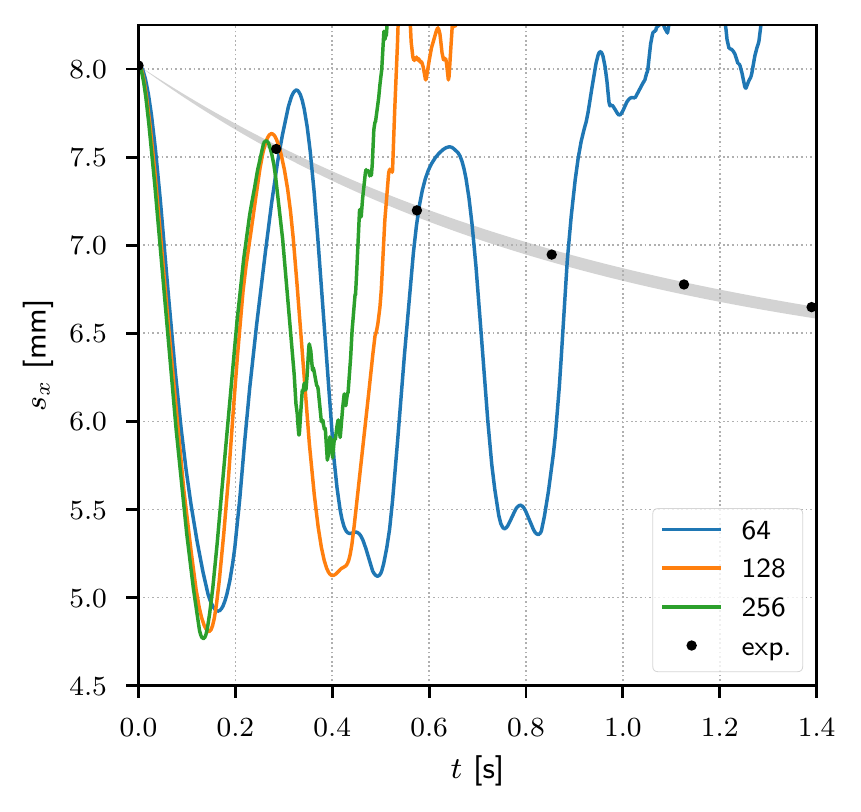}
        \caption{$\frac{L}{W} = 1.47$, old state}
        \label{plot:od-exp-sx-lw147-old}
    \end{subfigure}
    ~
    \begin{subfigure}[b]{0.48\textwidth}
        \includegraphics[width=\textwidth]{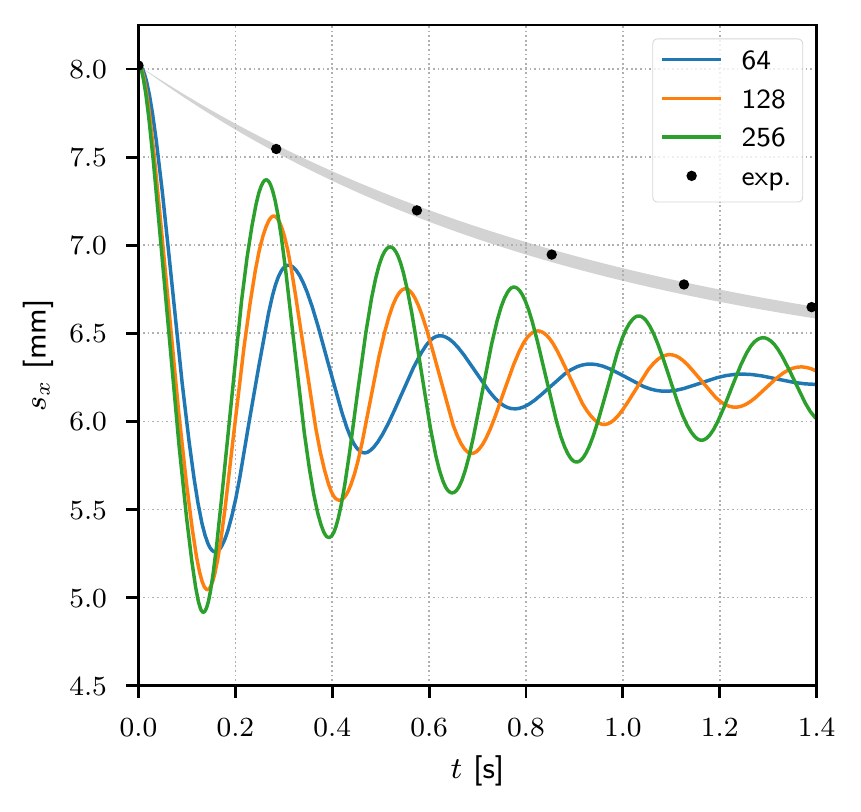}
        \caption{$\frac{L}{W} = 1.47$, new state}
        \label{plot:od-exp-sx-lw147-new}
    \end{subfigure}
    ~
    \begin{subfigure}[b]{0.48\textwidth}
        \includegraphics[width=\textwidth]{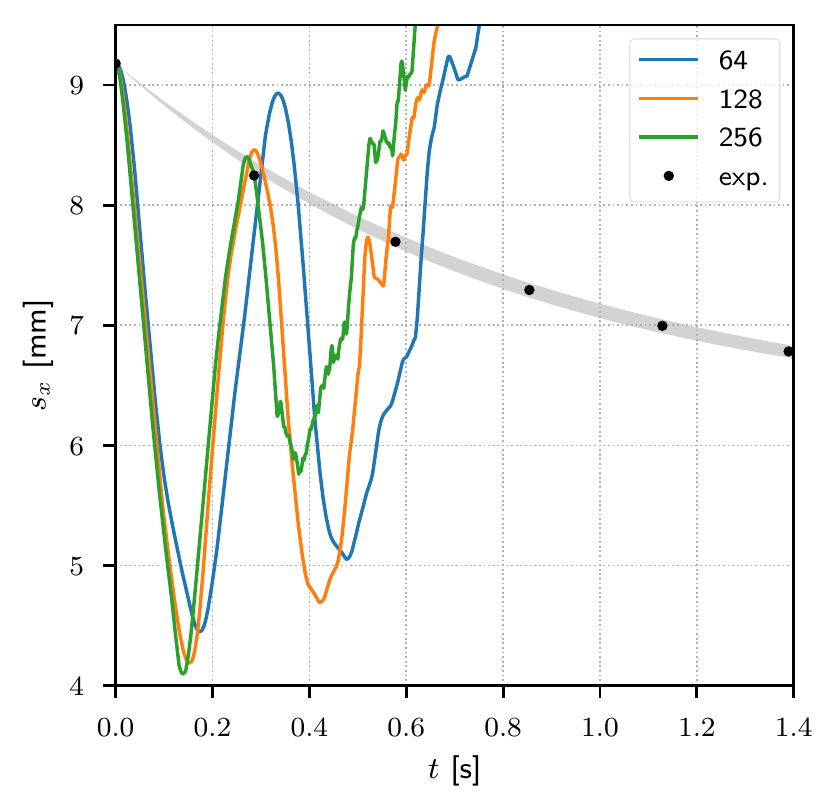}
        \caption{$\frac{L}{W} = 1.80$, old state}
        \label{plot:od-exp-sx-lw180-old}
    \end{subfigure}
    ~
    \begin{subfigure}[b]{0.48\textwidth}
        \includegraphics[width=\textwidth]{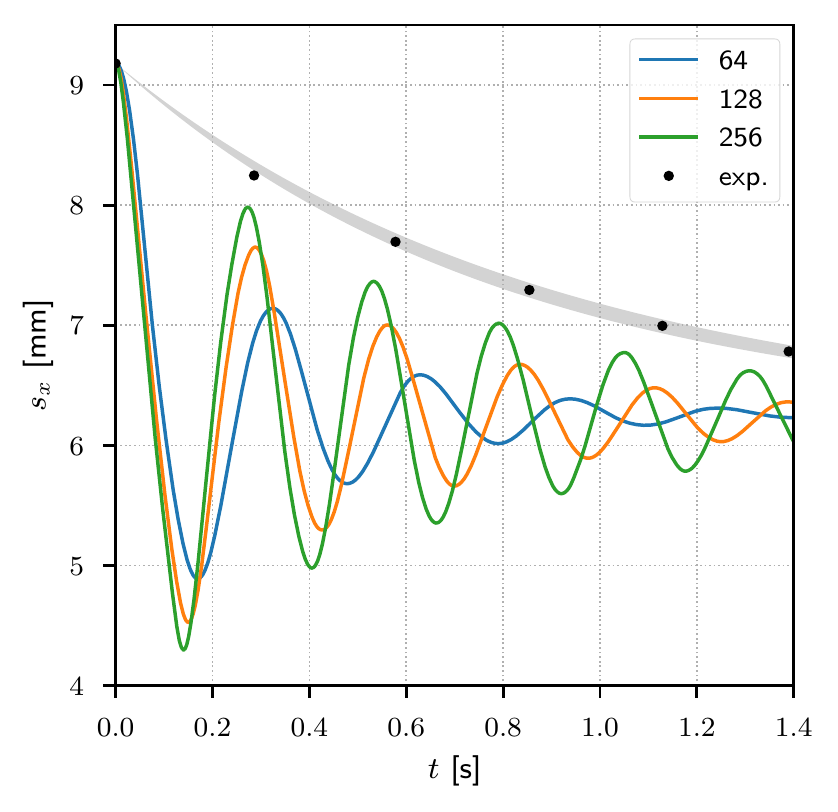}
        \caption{$\frac{L}{W} = 1.80$, new state}
        \label{plot:od-exp-sx-lw180-new}
    \end{subfigure}
    \caption{Temporal evolution of the $x$ semi-axis $\Semix$ for the oscillating droplet replicating the experimental setup described in \citep{Trinh1982}. The gray area
    depicts the decay envelope (see \cref{eq:decay-rate}) according to the experimentally
    measured $\gamma$, while the black dots mark measured oscillation peaks. Each plot shows the results for three different mesh resolutions $n_e$. The left column shows the results
    using the LENT configuration from \citep{Maric2015}, the right column the results obtained with the current configuration. Two semi-axes ratios $L/W$ have been simulated.}
    \label{fig:od-large-amplitude}
\end{figure}
In \cref{fig:od-large-amplitude}, the previous configuration of LENT \citep{Maric2015} and the one described in this publication are compared to
Trinh and Wang's experimental results given in section five of \citep{Trinh1982}. The
semi-axis $\Semix$ is evaluated according to \cref{eq:semi-axis-computation}. Except
for the lowest resolution $n_e=64$, the old configuration shows good agreement
of amplitude and period for the first peak at $t\approx0.28~\text{s}$. However, afterwards, the behavior is qualitatively similar to \cref{fig:oscillating-droplet-semiaxis-evolution}. Due to parasitic currents, perturbations accumulate in the front and feed back into the velocity field through surface tension.
Subsequently, the semi-axis evolution starts to severely deviate from the expected behavior during the second oscillation period. Between $t\approx0.35$~s and $t\approx0.5$~s, depending on resolution and
semi-axes ratio, the graphs no longer resemble a harmonic oscillation. With the modifications proposed here, however, the simulations yield the qualitatively expected
behavior. Agreement between the experimental oscillation period and the simulated one
is quite good with a relative difference of
$e_\text{rel}(L/W=1.47)\approx0.07$ and $e_\text{rel}(L/W=1.80)\approx0.05$ for $n_e=256$.
The amplitude decays noticeably faster in the simulation compared to the experiment.
This is related to the reconstruction operator \cref{eq:reconstruct} and its diminishing convergence, illustrated in \cref{subsec:result:forcerecon}. Another cause lies in the semi-implicit
surface tension model \cref{eq:semi-implicit-csf} as the second term is
effectively a diffusion term.
Improvement of the balanced discretization between the surface tension force and the pressure gradient on unstructured meshes by introducing an alternative field reconstruction operator in OpenFOAM, as well as the introduction of the new algorithm for the reconstruction of the Front are ongoing work. It is important to note, though, that the simulation results computed with the existing numerical method converge toward the experimental data with increasing mesh resolution. 

%% file: tables/curvatureErrors-exactDistance.tex
\begin{tabular}{lllrrrr}
\toprule
                   &              &     & $\overline{L_\infty}(e_{\MeanCurvature,\text{rel}})$ & $\overline{L_2}(e_{\MeanCurvature,\text{rel}})$ & $\sigma(L_\infty(e_{\MeanCurvature,\text{rel}}))$ & $\sigma(L_2(e_{\MeanCurvature,\text{rel}}))$ \\
model & interface & $n$ &                                           &                                     &                                        &                                   \\
\midrule
\multirow{14}{*}{\compactsphere{}} & \multirow{7}{*}{ellipsoid} & 16  & 1.17e+00 & 1.10e-01 & 1.20e-01  &  7.99e-03 \\
                                                   &&  -  & 0.83     & 1.97     & -         &  -        \\
                                                   && 32  & 6.58e-01 & 2.80e-02 & 9.97e-02  &  4.99e-04 \\
                                                   &&  -  & 1.19     & 2.05     & -         &  -        \\
                                                   && 64  & 2.88e-01 & 6.74e-03 & 2.42e-02  &  1.41e-04 \\
                                                   &&  -  & 1.16     & 1.79     & -         &  -        \\
                                                   && 128 & 1.29e-01 & 1.95e-03 & 4.34e-03  &  1.46e-05 \\
\cline{2-7}
                   & \multirow{7}{*}{sphere} & 16  &  9.37e-03 &  8.14e-04 &  5.16e-04 &  2.27e-05 \\
                                            &&  -  &  2.69     &  2.87     &  -        &  -        \\
                                            && 32  &  1.45e-03 &  1.11e-04 &  2.21e-05 &  5.06e-07 \\
                                            &&  -  &  2.32     &  2.62     &  -        &  -        \\
                                            && 64  &  2.91e-04 &  1.80e-05 &  1.67e-06 &  5.84e-08 \\
                                            &&  -  &  2.10     &  2.53     &  -        &  -        \\
                                            && 128 &  6.79e-05 &  3.12e-06 &  2.04e-07 &  5.10e-09 \\
\cline{1-7}
\cline{2-7}
    \multirow{14}{*}{\dgMarkerfield{}} & \multirow{7}{*}{ellipsoid} & 16  &  1.42e+01 &  8.19e-01 &  3.52e+00 &  6.78e-02 \\
                                                              &&  -  &  0.86     &  1.84     &  -        &  -        \\
                                                              && 32  &  7.85e+00 &  2.29e-01 &  8.49e-01 &  1.30e-02 \\
                                                              &&  -  &  0.10     &  1.77     &  -        &  -        \\
                                                              && 64  &  7.32e+00 &  6.72e-02 &  4.32e+00 &  7.56e-03 \\
                                                              &&  -  &  -0.59    &  0.14     &  -        &  -        \\
                                                              && 128 &  1.10e+01 &  6.09e-02 &  2.09e+00 &  1.25e-03 \\
\cline{2-7}
                   & \multirow{7}{*}{sphere} & 16  &  2.14e+00 &  1.72e-01 &  3.00e-01 &  2.19e-03 \\
                                            &&  -  &  -0.29    &  1.40     &  -        &  -        \\
                                            && 32  &  2.62e+00 &  6.50e-02 &  8.08e-01 &  2.32e-03 \\
                                            &&  -  &  -0.93    &  0.87     &  -        &  -        \\
                                            && 64  &  5.00e+00 &  3.56e-02 &  1.81e+00 &  2.02e-03 \\
                                            &&  -  &  -0.63    &  0.27     &  -        &  -        \\
                                            && 128 &  7.74e+00 &  2.96e-02 &  4.48e+00 &  1.44e-03 \\
\bottomrule
\end{tabular}

%% file: tables/curvatureErrors-frontDistance.tex
\begin{tabular}{lllrrrr}
\toprule
                   &              &     & $\overline{L_\infty}(e_{\MeanCurvature,\text{rel}})$ & $\overline{L_2}(e_{\MeanCurvature,\text{rel}})$ & $\sigma(L_\infty(e_{\MeanCurvature,\text{rel}}))$ & $\sigma(L_2(e_{\MeanCurvature,\text{rel}}))$ \\
model & interface & $n$ &                                 &                            &                              &                         \\
\midrule
\multirow{14}{*}{\compactsphere{}} & \multirow{7}{*}{ellipsoid} & 16  &  1.12e+00 &  1.09e-01 &  1.59e-01 &  9.13e-03 \\
                                                   &&  -  &  0.76     &  1.95     &  -        &  -        \\
                                                   && 32  &  6.60e-01 &  2.83e-02 &  8.38e-02 &  5.74e-04 \\
                                                   &&  -  &  1.13     &  2.03     &  -        &  -        \\
                                                   && 64  &  3.02e-01 &  6.91e-03 &  2.68e-02 &  1.34e-04 \\
                                                   &&  -  &  1.22     &  1.80     &  -        &  -        \\
                                                   && 128 &  1.30e-01 &  1.98e-03 &  2.86e-03 &  9.53e-06 \\
\cline{2-7}
&\multirow{7}{*}{sphere} & 16  &  8.45e-03 &  1.07e-03 &  5.67e-04 &  5.39e-05 \\
                        &&  -  &  0.46     &  1.50     &  -        &  -        \\
                        && 32  &  6.16e-03 &  3.79e-04 &  4.96e-04 &  8.87e-06 \\
                        &&  -  &  -0.01    &  0.87     &  -        &  -        \\
                        && 64  &  6.20e-03 &  2.05e-04 &  1.12e-03 &  3.73e-06 \\
                        &&  -  &  -0.29    &  0.83     &  -        &  -        \\
                        && 128 &  7.59e-03 &  1.15e-04 &  1.48e-03 &  2.44e-06 \\
\cline{1-7}
\cline{2-7}
    \multirow{14}{*}{\dgMarkerfield{}} & \multirow{7}{*}{ellipsoid} & 16  &  1.31e+01 &  7.78e-01 &  4.23e+00 &  7.97e-02 \\
                                                               &&  -  &  0.78     &  1.77     &  -        &  -        \\
                                                               && 32  &  7.61e+00 &  2.28e-01 &  9.29e-01 &  1.22e-02 \\
                                                               &&  -  &  0.1      &  1.77     &  -        &  -        \\
                                                               && 64  &  7.08e+00 &  6.69e-02 &  4.21e+00 &  7.75e-03 \\
                                                               &&  -  &  -0.69    &  0.12     &  -        &  -        \\
                                                               && 128 &  1.14e+01 &  6.14e-02 &  4.39e+00 &  2.78e-03 \\
\cline{2-7}
                   & \multirow{7}{*}{sphere} & 16  &  2.23e+00 &  1.73e-01 &  3.02e-01 &  2.54e-03 \\
                                            &&  -  &  -0.53    &  1.37     &  -        &  -        \\
                                            && 32  &  3.22e+00 &  6.68e-02 &  8.69e-01 &  2.49e-03 \\
                                            &&  -  &  -0.44    &  0.93     &  -        &  -        \\
                                            && 64  &  4.36e+00 &  3.51e-02 &  1.62e+00 &  1.94e-03 \\
                                            &&  -  &  -0.73    &  0.26     &  -        &  -        \\
                                            && 128 &  7.23e+00 &  2.93e-02 &  4.22e+00 &  1.35e-03 \\
\bottomrule
\end{tabular}

%% file: tables/sd-piso-iterations.tex
\begin{tabular}{l|rrrr}
    \toprule
    $n_e$ & \multicolumn{4}{c}{Laplace number} \\
    & 120 & 1200 & 12000 & $\infty$ \\
    \hline
    16 & 15 & 9 & 7 & 1 \\
    64 & 19 & 11 & 8 & 1 \\
    \bottomrule
\end{tabular}

%% file: tables/stationary_droplet_exact_curvature_PISO_LSC.tex
\begin{tabular}{llrrrrr}
    \toprule
    && \multicolumn{2}{c|}{$\Linf{\Norm{\Velocity(t=\Delta t)}}$}
    & \multicolumn{2}{c|}{$\Linf{\Norm{\Velocity(t=t_\text{end})}}$} 
    & $t_\text{end}$\\
    && \multicolumn{1}{c|}{PISO} & \multicolumn{1}{c|}{\Saample}
    & \multicolumn{1}{c|}{PISO} & \multicolumn{1}{c|}{\Saample} & \\
    $La$ & $n_e$ &&&&& \\
    \midrule
    \multirow{4}{*}{120} & 16 & 3.90e-06 & 1.47e-14 & 3.32e-14 & 1.14e-14 & 7.8 \\
                         & 32 & 7.40e-06 & 7.74e-15 & 1.25e-14 & 3.74e-15& 7.8 \\
                         & 64 & 1.49e-05 & 1.77e-14 & 2.17e-14 & 3.40e-14& 7.8 \\
                         & 128 & 2.37e-05 & 1.29e-14 & 8.55e-11 & 3.61e-14& 0.3 \\
                         \cline{2-7}
    \multirow{4}{*}{1200} & 16  & 9.29e-08 & 1.06e-14 & 7.75e-16 & 1.33e-15 & 24.8 \\
                          & 32  & 2.25e-07 & 2.70e-14 & 6.32e-15 & 6.62e-16 & 24.8 \\
                          & 64  & 6.10e-07 & 1.56e-14 & 5.47e-14 & 9.84e-16 & 8.0 \\
                          & 128 & 1.39e-06 & 1.61e-14 & 2.53e-11 & 1.72e-14 & 0.3 \\
                          \cline{2-7}
    \multirow{4}{*}{12000} & 16  & 1.27e-09 & 4.26e-14 & 1.39e-14 & 1.72e-14 & 78.4 \\
                           & 32  & 3.38e-09 & 4.85e-14 & 1.59e-15 & 1.37e-15 & 50.0 \\
                           & 64  & 1.07e-08 & 1.74e-14 & 2.67e-13 & 2.66e-14 & 8.0 \\
                           & 128 & 2.99e-08 & 1.94e-14 & 5.81e-12 & 2.77e-14 & 0.3 \\
                           \cline{2-7}
    \multirow{4}{*}{$\infty$} & 16  & 2.60e-15 & 7.29e-14 & 2.03e-13 & 1.75e-14 & 100 \\
                              & 32  & 2.97e-15 & 2.34e-14 & 1.56e-13 & 1.25e-13 & 35.0 \\
                              & 64  & 2.73e-15 & 2.18e-14 & 1.58e-13 & 7.54e-14 & 8.0 \\
                              & 128 & 2.70e-15 & 1.58e-14 & 2.19e-14 & 1.20e-14 & 0.3 \\
    \bottomrule
\end{tabular}

%% file: sections/conclusions.tex
The proposed SAAMPLE algorithm together with the improvements in the curvature approximation, phase indicator approximation and implicit surface tension modeling significantly increases the numerical robustness of the unstructured LENT hybrid Level Set / Front Tracking method when simulating surface-tension driven flows, when compared to both the previous publication \citep{Maric2015} and contemporary Level Set and VOF methods on structured meshes. For the experimental case reported in \cref{subsection:od-large-amplitude}, overdamping of the solution is still present. However, the solution converges to the experimental observation with increased mesh resolution.

We have found that the field reconstruction from scalar values on unstructured meshes in OpenFOAM diverges for fields that are not at least $C^1$. This behavior of the reconstruction operator has not been reported so far in the literature, and it is crucial for the segregated equation coupling in OpenFOAM for multiphase flows. The field reconstruction is also used for combustion, spray simulations, electromagnetic simulations and heat transfer (weak compressibility), so the findings reported in \cref{subsec:num:forcerecon} might be of significant importance for those applications as well.

Additionally, the length scale of the reconstructed Front should be connected with the length scale of the Eulerian mesh by developing a new Front reconstruction algorithm on unstructured meshes which does not construct the connectivity of Front elements. The absence of connectivity between the Front elements will make the parallelization of the method, using the message passing parallel programming model, more straightforward and will enable us to more accurately tackle physical problems such as the one in \cref{subsection:od-large-amplitude}, by allowing much higher mesh resolutions.

Improvements of the field and Front reconstruction algorithms are left as future work.

%% file: sections/acknowledgements.tex
We kindly acknowledge the financial support by the German Research Foundation (DFG) within the Initiation of International Collaboration "Hybrid Level Set / Front Tracking methods for simulating multiphase flows in geometrically complex systems", MA 8465/1-1.

Calculations for this research were conducted on the Lichtenberg high performance computer of the TU Darmstadt.

The authors are very grateful to Dr.\ Damir Juric (LIMSI institute, CNRS) for his advice regarding the compact curvature calculation.